\documentclass[10pt,journal,finalsubmission,compsoc,twoside,final]{IEEEtran} 

\usepackage[nocompress]{cite}
\usepackage{graphicx}
\graphicspath{{./}}
\DeclareGraphicsExtensions{.pdf}
\usepackage[cmex10]{amsmath}
\usepackage{array}
\usepackage{mdwmath}
\usepackage{mdwtab}
\usepackage{caption}
\usepackage{fixltx2e}
\usepackage{url}

\begin{document}


\title{\emph{Shape}: A 3D Modeling Tool for Astrophysics}

\author{Wolfgang~Steffen, Nicholas~Koning, Stephan~Wenger, Christophe~Morisset, and Marcus~Magnor,~\IEEEmembership{Member,~IEEE}\IEEEcompsocitemizethanks{
\IEEEcompsocthanksitem Wolfgang Steffen is with the Instituto de Astronom\'{\i}a, Universidad Nacional Aut\'onoma de M\'exico, Ensenada, B.C., Mexico.\protect\\ E-mail: wsteffen@astrosen.unam.mx
\IEEEcompsocthanksitem Nicholas Koning is with the Space Astronomy Laboratory, University of Calgary, Calgary, Canada.\protect\\ E-mail: nkoning@iras.ucalgary.ca
\IEEEcompsocthanksitem Stephan Wenger and Marcus Magnor are with the Computer Graphics Lab, TU Braunschweig, Germany.\protect\\ E-mail: wenger@cg.tu-bs.de, magnor@cg.tu-bs.de
\IEEEcompsocthanksitem Christophe Morisset is with the Instituto de Astronom\'{\i}a, Universidad Nacional Aut\'onoma de M\'exico, Ciudad Universitaria, D.F., Mexico.\protect\\ E-mail: Chris.Morisset@gmail.com}}
\markboth{IEEE Transactions on Visualization and Computer Graphics,~Vol.~6, No.~1, January~2007}{Steffen \MakeLowercase{\textit{et al.}}: \emph{Shape}: A 3D Modeling Tool for Astrophysics}

\IEEEcompsoctitleabstractindextext{
\begin{abstract}
We present a flexible interactive 3D morpho-kinematical modeling application for astrophysics. Compared to other systems, our application reduces the restrictions on the physical assumptions, data type and amount that is required for a reconstruction of an object's morphology. It is one of the first publicly available tools to apply interactive graphics to astrophysical modeling. The tool allows astrophysicists to provide a-priori knowledge about the object by interactively defining 3D structural elements. By direct comparison of model prediction with observational data, model parameters can then be automatically optimized to fit the observation. The tool has already been successfully used in a number of astrophysical research projects.
\end{abstract}}

\maketitle
\IEEEdisplaynotcompsoctitleabstractindextext
\IEEEpeerreviewmaketitle


\section{Introduction}

The interpretation of astrophysical data often depends strongly
on the knowledge of depth information along the line of sight.
In most cases, however, this is the least well known information.
That is true for the distance and especially for
the position of substructure {\em within} an object.
The development of effective methods for the reconstruction of
the 3D structure of astrophysical objects is therefore an issue of growing
importance in astronomy.
Photographic images only provide a two-dimensional integration
of the emission and absorption along the line of sight. The depth information is
therefore flattened. Sometimes, symmetry properties combined with a favorable
orientation of an object
provide sufficient information to visually deduce what the structure
must be. This can be the case for planetary nebulae and has been used
to automatically reconstruct the 3D structure
(Leahy~\cite{Leahy91:DEA}, Magnor et al.~\cite{Magnor04:CIVRPN,Magnor05:RVPN},
Lin\c{t}u et~al.~\cite{Lintu07:MDGDPN,Lintu07:3DEAPN}).
If no such
symmetries are present, then the depth information must come from
other types of information, which usually depend on a fundamental physical model
for the object class that is considered. This information could be
the velocity field, e.g. in a radially expanding nebula a mapping between
velocity and position exists.
However, for some objects -- such as turbulent interstellar clouds --
such a mapping is not possible.

Much of observational astrophysics research involves physical modeling
with limited constraints to deduce physical properties of the observed
objects. Astrophysicists measure a limited
number of physical properties of an object, via
electromagnetic waves, to which a physical model of the
phenomenon is then fitted. Such models are usually not unique.

Most of the astrophysical modeling effort tends to gravitate towards massive
parallel supercomputing for dynamical simulations.
Analysis and visualization of such simulations are done separately and often are
complex and computationally intense processes themselves~\cite{Henney09}.
While such simulations produce insight into generic astrophysical
processes, they are rarely suitable for elucidating the properties and
structure of particular objects.

Knowing the properties of individual objects
is essential when a single object class, e.g. planetary nebulae, shows
a large variety of presentations in images.
A serious difficulty for the modeling of particular objects is our fixed vantage
point on Earth which restricts all observations to be along a single direction
(up to the parallax provided by the Earth's orbit, which is negligible 
for typical distances to astronomical nebulae).
This is in strong contrast to, e.g., medical imaging
where 3D information is recovered from observations from multiple directions
around the subject.
For any astronomical object beyond the solar system, we are able to
observe only one 2D projection of its actual 3D volumetric shape.
For the correct physical interpretation of observational data, information
about the object's actual 3D shape has to be
available~\cite{Lucy74:ITR,Bremer95:TDI,Palmer94:DAG,Leahy94:IDA,Volk93:DPN}.
Obtaining new structural information and insight on particular
objects is the main purpose of the application that we present in this paper.
Our approach to modeling individual objects is very different
from previous methods.
The application that we present (called {\it Shape}) becomes essential
when automatic reconstruction methods fail because theoretical or
observational constraints are insufficient.
A lack of constraints for an automatic reconstruction is at least partially
compensated by scientific user judgement.
Often the available constraints are sufficient to test one or more hypotheses about
the structure of an object.

In such cases, rather than a reconstructive, a constructive morpho-kinematical modeling
approach is more suitable to reconstruct the structure and velocity field.
The term morpho-kinematical is applied to modeling that involves only structural
(morphological) and velocity (kinematic) information.
This is in contrast to dynamical
simulations, which include the effects of forces and temporal evolution from a set
of simpler initial and boundary conditions.
In general, the outcome of dynamical simulations is not predictable in detail and
very hard to tune to a specific object.

Conventional morpho-kinematical modeling uses hard-coded mathematical descriptions
of the objects, processes and boundary conditions. Therefore, the user needs
at least basic programming skills in the particular language
of the code.
Modern 3D modeling software of the graphics industry shows that such modeling
can be done effectively without user programming intervention. Although such software
can visualize gas-dynamical processes, it is inefficient and its usefulness for
astrophysical processes is limited.
The general workflow of such systems seemed, however,
very suitable for modeling particular astrophysical objects~\cite{Steffen06}.

\begin{figure}[t]
\centering
\includegraphics[width=0.45\textwidth]{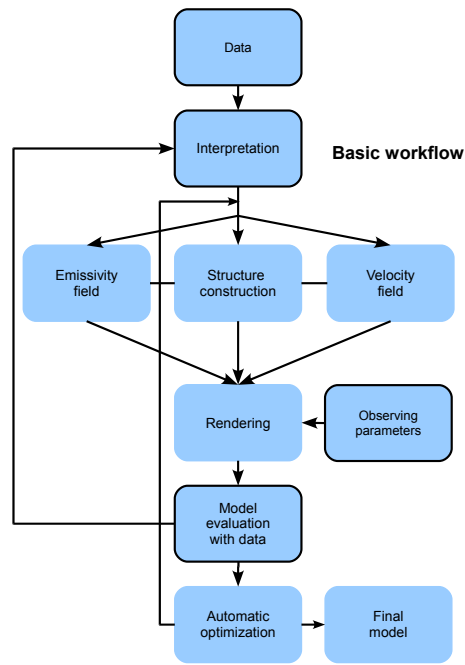}
\caption{The \emph{Shape} modeling tool that we present here
provides a unified framework for the whole modeling and visualization workflow.
The user employs his physical knowledge to construct an initial model
which can be visualized and compared to observational data in several ways,
allowing for easy interactive and iterative refinement of the model.
When all necessary physical information is reflected in the model,
its parameters can be automatically optimized,
minimizing the difference between the model and the observational data.
The final model can then be used to generate various types of graphical output.}
\label{workflow}
\end{figure}

Following the technique of modern interactive 3D graphics systems,
we have developed {\it Shape} with specialized functionality for
interactive astrophysical modeling on single desktop or laptop computers.
The primary purpose of {\it Shape} is to interactively generate 3D models.
However, in contrast to
conventional astrophysical modeling tools, it integrates the visualization
and analysis of the model into
the same system. Direct access to the model data at any stage of the modeling
process allows for effective comparison with the observed data in a variety of ways
within the feedback loop of the iterative workflow (Figure~\ref{workflow}).

In artistic work on astronomical topics, commercial tools like Maya or
3D Studio Max are frequently being
employed~\cite{Yu05:DFD,Davis05:STA,Matthews05:DDF}.
Professional animation tools are designed to assist in
creating realistic 3D scenes of familiar environments.
Unfortunately, when used for scientific work they display serious
shortcomings.
Especially volume rendering with mesh structures and particle systems
are very different from the physical correct radiation transfer needed for
reliable interpretation of astrophysical phenomena.
A qualitative and quantitative comparison of such models with real objects
is not possible.
The key problem preventing their use in astrophysics research is the
inability to produce the type of renderings that are comparable to
the observations obtained with telescopes and other scientific instrumentation like
spectrographs (e.g., for Doppler-shift measurements).

With \emph{Shape} we remedy most of the shortcomings of
previous astrophysical reconstruction systems by applying the powerful
structure modeling techniques of commercial animation suites, while
adding the information output and processing systems that
are necessary for astrophysical research applications.
We go beyond the current commercial rendering techniques by using physically
more accurate modeling of the radiation transfer from the sources
to the observer.

In this paper we first comment on previous related work in Section
\ref{related}, on the type of observational data that are used for this
work in Section \ref{data} and then introduce the \emph{Shape} system.
In Section \ref{results} we show three examples of previously published
research applications of \emph{Shape}, before giving an outlook on
future developments and our conclusions in Sections \ref{future} and
\ref{conclusion}, respectively.

\section{Related Work}
\label{related}

Many different approaches have historically been used to simplify the modeling of astrophysical objects.
Besides the general reconstruction techniques applying to a wider range of volumetric objects (cf.~\cite{Ihrke08}),
specialized approaches have been developed that exploit the peculiarities of the astrophysical case.
Among them are automatic methods that make use of Doppler shift measurements (cf. the section on observational data) or symmetry assumptions as well as user-driven modeling systems.

\subsection{Automatic Reconstruction Methods}
\label{automatic}

\begin{figure}[t]
\centering
\includegraphics[width=0.45\textwidth]{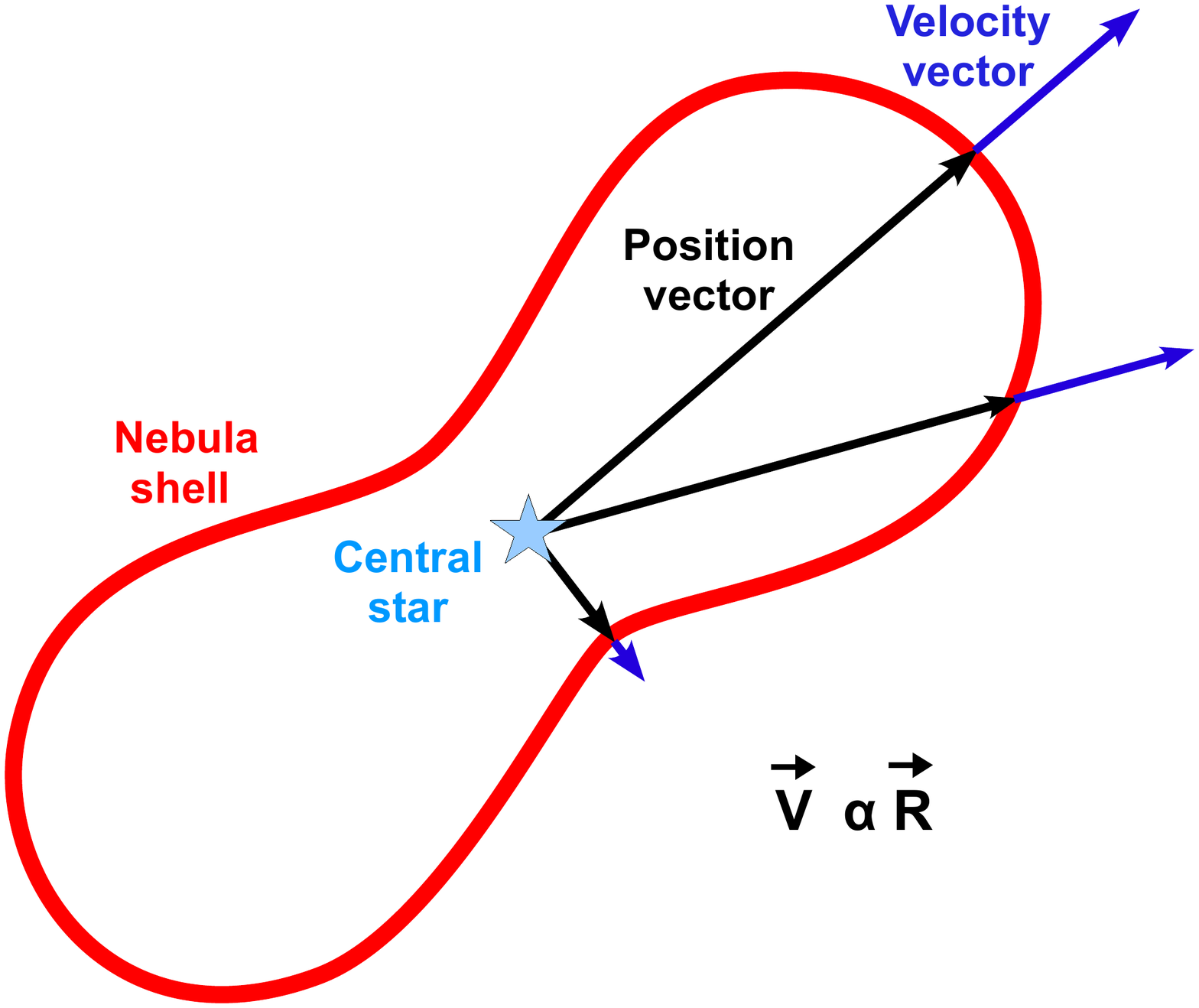}
\includegraphics[width=0.45\textwidth]{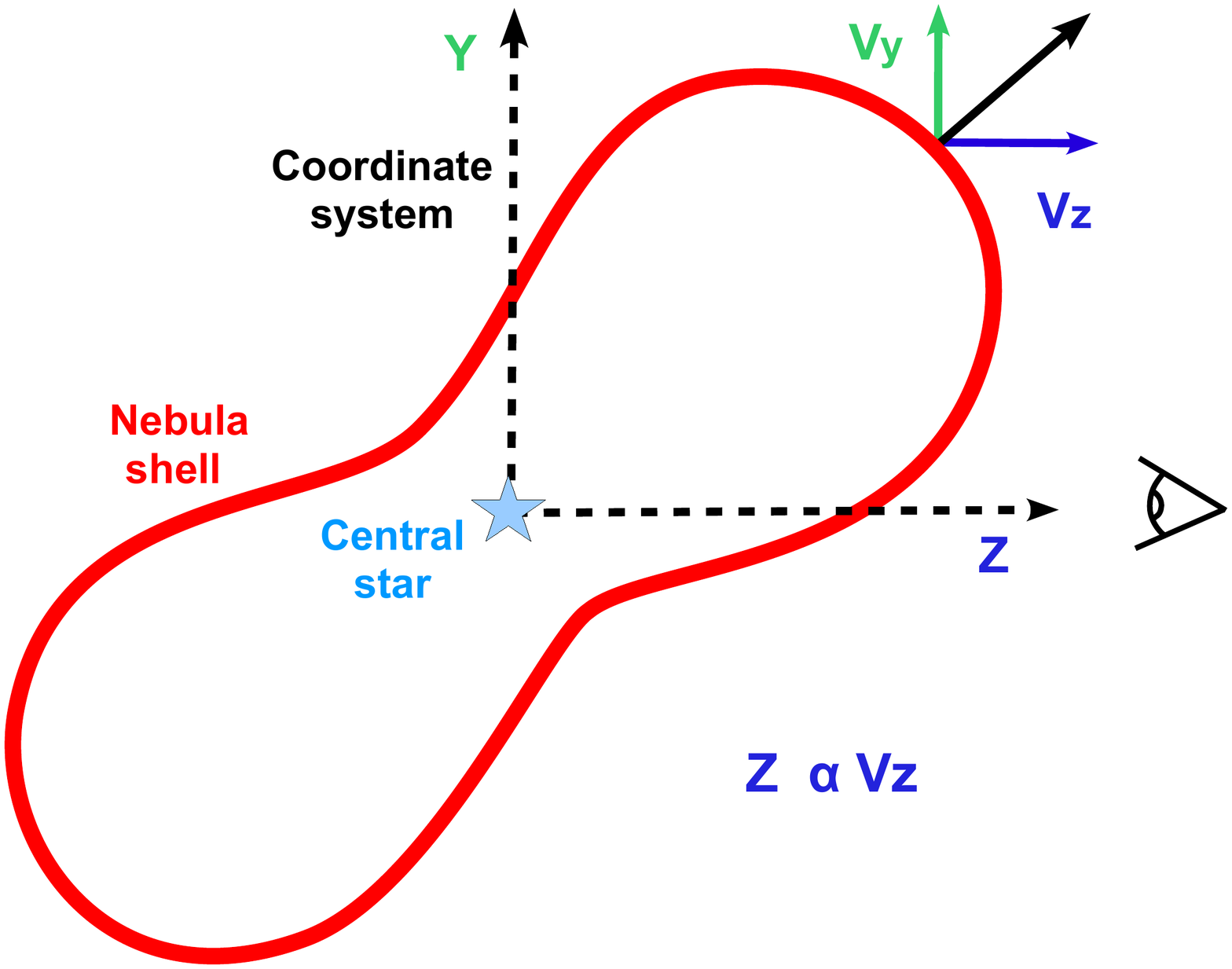}
\caption{The top panel shows the outline of a bipolar nebula with
position and velocity vectors marked.
Most 3D reconstruction systems for expanding astrophysical nebulae
are based on the assumption that the velocity vector is proportional to the
position vector.
The Doppler-shift of a spectral line due to the velocity component along the line of
sight (${\rm v_z}$, bottom panel) provides a direct mapping to the position
along the line of sight (${\rm z}$).
The unknown constant of proportionality can often be determined from
additional information, e.g. symmetries that are inherent in the object structure.
}
\label{doppler}
\end{figure}

Many extended astrophysical objects show a strong correlation between the velocity
and position of the emitting gas relative to some local reference point, which often is
the center of a star or stellar remnant.
For example, material that has been ejected ballistically with different velocities
from the same source (e.g. by an explosion) will naturally evolve such that
-- after a time period that is long enough compared to the duration of the ejection process --
the faster regions have moved farther away from the source and an approximately linear dependence
between velocity and distance from the source is established (see Figure \ref{doppler}, top).

Doppler shift methods like the one by Sabbadin et al.~\cite{sabbadin,tlepn}
make use of this correlation in order to derive depth information from Doppler-shift data.
If the assumption of linear dependence between the position and velocity vector holds,
a linear mapping exists between the Doppler--shift (i.e. velocity along the line of sight)
and the position along the same direction (see Figure \ref{doppler}, bottom).
In this case the resulting models are accurate within the limits of the accuracy of
the Earth-bound observational data.
Unfortunately, many objects contain several different kinematic subsystems which may have
different relations between velocity and position.
Some also show complex interactions with their local environment which may further
complicate the velocity law~\cite{WSGGS2009}.
Furthermore, these methods require an almost complete coverage of the object with
regularly spaced observations of the Doppler-shift, which require special observing programs.
Such homogeneous data sets are rarely available.

Other algorithms that are more oriented towards high visual quality than physical accuracy of
the results are based on symmetry constraints
(Magnor et al.~\cite{Magnor04:CIVRPN,Magnor05:RVPN},
Lin\c{t}u et~al.~\cite{Lintu07:MDGDPN,Lintu07:3DEAPN}, Wenger et al.~\cite{Wenger09}).
Many astronomical nebulae show an inherent spherical or axial symmetry due
to their evolution from more or less symmetrical sources.
This symmetry assumption may be used to reconstruct the missing spatial
dimension~\cite{Leahy91:DEA}.

In many cases, however, these simple symmetries are disturbed by statistical effects or
external influences, and more complex symmetries as well as turbulent structures arise
which cannot be modeled in such a generic way.

\begin{figure}[t]
\centering
\includegraphics[width=0.5\textwidth]{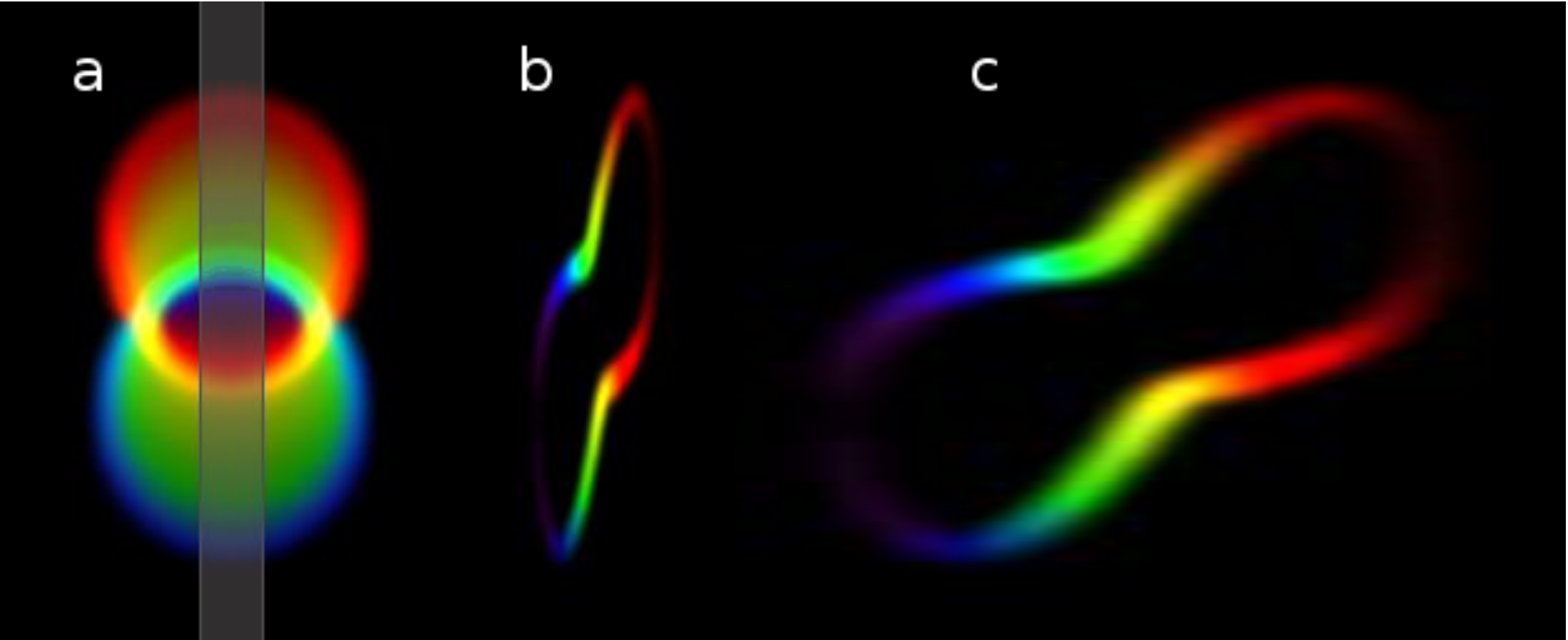}
\caption{Panel ``a'' shows the projected synthetic image of a bipolar nebula comparable to
the one outlined in Figure~(\ref{doppler}) with a similar orientation to the observer.
The color coding is according to the velocity along
the line of sight (blue is approaching and red is receding).
The marked synthetic spectrograph slit was used to generate the
P--V diagram in panel ``b''.
Since the velocity field has been assumed homologous,
the P--V diagram can be stretched along
the horizontal velocity axis such that its outline corresponds to
a cut through the object along the line of sight (panel ``c'').
}
\label{image_pv}
\end{figure}

\subsection{User-Driven Systems}

The above-mentioned shortcomings of fully automatic reconstruction approaches may be avoided
by resorting to interactive modeling techniques.
The interactive creation of a model aiming to reproduce a given single image is a common task,
but most existing solutions are not well suited for the modeling of emissive transparent objects
that are prevalent in astronomy
and do not allow for representation of velocity information and spectral data.
Among the tools that most closely reflect our modeling approach are
the interactive approaches of Debevec et al.~\cite{Debevec96},
Fran\c{c}ois and Medioni~\cite{Francois01} and Zhang et al.~\cite{Zhang01},
all of which expect some kind of user-specified coarse geometry or a set of
user-defined geometry constraints
which is then automatically converted into a full three-dimensional model that
best fits the provided image under the given constraints.
The idea of automatic optimization of a parameterized (deformable or ``morphable'') model
has been successfully employed in the works of Montagnat and Delingette~\cite{Montagnat98} and Romdhani and Vetter~\cite{Romdhani03}.

Some entirely manual modeling aids have been developed for the astrophysical use case,
in which an astronomer defines a model representing his or her high-level knowledge about the object in question.
This process will usually involve iterations of modeling, comparison of the rendered model
to actual observational data (especially spectral data), and refinement of the model,
until all observational facts can be explained by the physically plausible model.

An early tool for the rendering part of the astrophysical modeling process was our earlier
work~\cite{Holloway96}
which was able to reproduce many standard forms of observational data from a given model.
Similar codes have been used
by Santander-Garc\'{\i}a et al.~\cite{Santander04} and Hajian et al.~\cite{Hajian07}.
The model itself, however, still had to be hard-coded into the program, making
the modeling part inherently cumbersome.
Steffen and L\'opez~\cite{Steffen06}
later incorporated their spectral renderer into a commercial modeling system as a plugin.
This simplified the modeling process to a large extent, but performance and usability
were still far from the quality of an integrated modeling and rendering system.

The only astrophysics-suited tool that we know of which follows a paradigm similar
to that of our earlier work
is \emph{Hydra} which is under development at MIT~\cite{Hydra08,Hydra}.
Contrary to \emph{Shape}, \emph{Hydra} focuses on X-ray data instead of
the near-visible wavelengths which are suitable for realistic visualizations
aimed at public media like planetaria~\cite{Magnor10}. 
Also, the tool does not provide an interactive modeling system,
but models are defined using scripting and Constructive Solid Geometry (CSG).

\begin{figure*}[t]
\centering
\includegraphics[width=\textwidth]{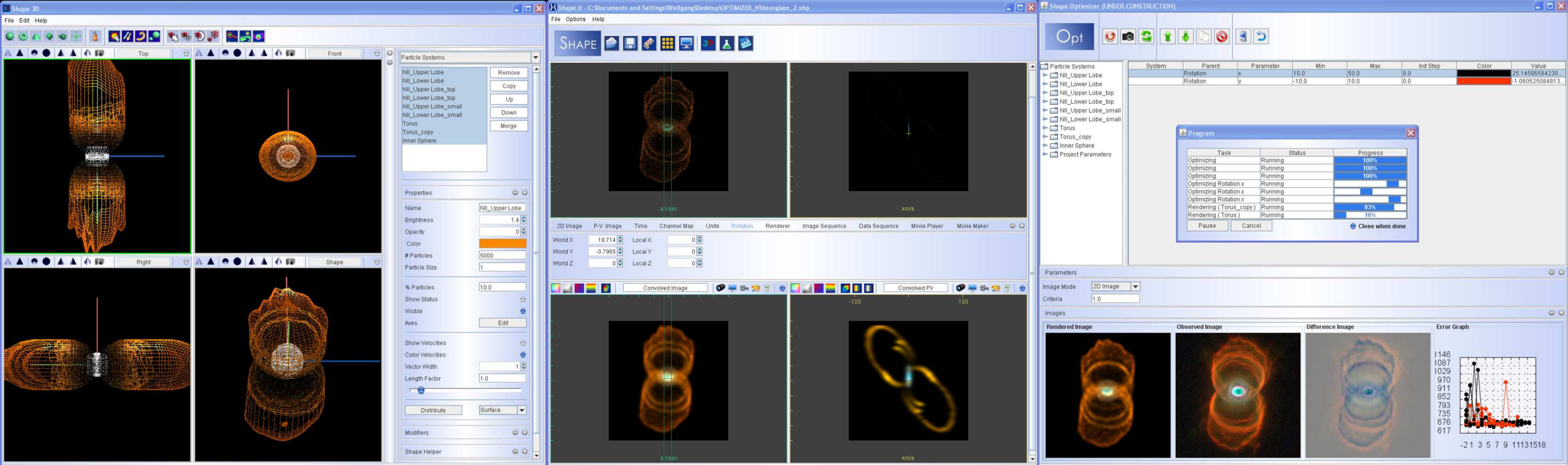}
\caption{The \emph{Shape} graphical user interface contains several modules
dedicated to different tasks.
Here, the interactive 3D modeling window,
the screen for rendering and comparison to observational data
and the control panel for automatic optimization
are shown, respectively, from left to right.
}
\label{shape_gui}
\end{figure*}

\section{Observational Data in Astronomy}
\label{data}

Many astrophysical objects (e.g. stars or emission nebulae) contain large amounts of ionized gas.
This ionized gas (or plasma) emits light at a number of well-known wavelengths, depending on
the chemical composition of the plasma.
Using filters for selected wavelengths, two-dimensional projections of
the spatial distributions of different ions within such an object can be measured
as photographs.
While the spatial resolution of earth-bound measurement devices is limited by
the perturbing effects of the atmosphere,
the advent of space telescopes such as the \emph{Hubble Space Telescope} (HST) made
high resolution imagery with high sensitivity available for a large number of objects.

Another key source of astronomical data for 3D reconstructions are \emph{Doppler-shift measurements}.
The underlying physical phenomenon is the Doppler-effect which causes
the wavelength of light to change when the emitter and the receiver have a
nonzero relative velocity towards each other.
With high-resolution spectrographs, it is possible
to measure the shift of any observed spectral line
with respect to its known reference wavelength, and therefore the relative velocity of the emitter
along the line of sight.
Because many complex astrophysical objects also have complex internal velocity fields,
the resulting distortions of the spectral lines are an important additional source of information.

To capture spatially varying wavelength information,
usually a narrow slit is used to select only a narrow, but long region of the object.
The light is then dispersed in wavelength along the direction perpendicular to the slit.
The result is recorded on an image detector and leads to a \emph{position-velocity} (P-V) \emph{diagram}.
The image intensity is the object intensity as a function of position along the slit
and wavelength (Figure~\ref{image_pv}).

\section{The \emph{Shape} System}
\label{shape}

The motivation for the development of \emph{Shape} was to
be able to reconstruct the 3D structure of astrophysical objects
based on data sets that do not allow an automatic reconstruction.
This may be because of a lack of sufficiently detailed data or
because the basic assumptions for the reconstruction algorithm
are not met by the object.
Therefore, a user-driven interactive approach was adopted.
The minimum general requirements that such a system should have
are those present in previous systems, which basically are:
\begin{itemize}
\item Tools to define a spatial emissivity and velocity field.
\item Establish a mapping between Doppler--shift and position.
\item Produce output that can be compared with observed images and spectra.
\item \emph{Shape}: The tools to define the spatial structure and velocity field
should be interactive
\end{itemize}
The first three requirements are for previously existing systems.
The last item, in italics, is the fundamental additional distinctive
requirement for \emph{Shape}.

To fulfill these four requirements, three major \emph{Shape} modules
are typically used when constructing a model to fit given observational data.
The workflow is illustrated in Figure (\ref{workflow}).
First, the \emph{interactive modeling} environment is used
to specify the user's assumptions about the object's geometry, velocity and emissivity.
This model is then input to the rendering module
which emulates how the modeled object would be observed
through a physical measurement device such as a telescope with a spectrograph.
Visualizations have been designed such that they are an aid to
obtain new physical insight and/or be comparable to actual astronomical
data.
The user then interactively refines the model until it qualitatively
fits the observational data.
In a final \emph{automatic optimization} step,
the numerical parameters of the model are varied
to also quantitatively fit the model in the best possible way.

The interactive modeling approach makes it important to keep the system highly
functional on single computers.
Parallelization therefore focuses on the
application of multi-core threading and, in the future, on parallel simulation
and rendering on graphics processing units (GPUs).

\subsection{{\em Java} Implementation}

For the choice of {\em Java} as the programming language for
{\em Shape} the following three initial criteria were decisive.
Astrophysicists use a range of operating
systems, mainly \emph{Linux}, \emph{Mac OS X} and \emph{Windows}.
Therefore, first, portability of the software and file formats was essential.
Second, since the software was to be developed over a considerable time and updated regularly,
{\em Java Webstart} seemed an excellent way to handle user-friendly updates.
Third, all essential Application Programming Interfaces (APIs) are
readily available.

In practice, however, there are a number of compatibility issues, especially with \emph{Mac OS X}.
The necessary {\em Java} software has become available only recently and only
for 64-bit systems.
Minor non-essential compatibility issues remain with \emph{Linux} and 64-bit \emph{Mac OS X} systems,
which can be expected to disappear as these operating systems and {\em Shape} evolve.

{\em Shape} requires four basic \emph{Java} components: the \emph{Java Virtual Machine} (JVM), {\em Java3D},
 \emph{Java Advanced Imaging} (\emph{JAI}) and \emph{Java Media Framework} (\emph{JMF}).
The 3D modeling environment is implemented with \emph{Java3D}, and \emph{JAI} has been used for some image
processing tasks. \emph{JMF} allows the encoding of animated image sequences into common movie formats.

Depending on the selected algorithm and the complexity of a model,
simulation times may range between fractions of a second and about one or two minutes
on a current Intel Core Duo processor with 2~GB~RAM,
but they usually stay below a few seconds for the most common cases with typical 256 pixel image sizes.

\subsection{Interactive Modeling}

The modeling system that we present here has to solve two key problems
that are common in existing astronomical 3D reconstruction tools.
Since the spatial and spectral resolution of current observational techniques
reveal a great wealth of irregular detail in gaseous nebulae,
simple mathematical descriptions of the structures have become impractical
as the number of parameters to handle can go into the hundreds.
We want our system to be able to model the complex multipolar, knotty or filamentary
structures that are commonly observed in gaseous nebulae.
Also, most earlier solutions have been based on ``off-line'' modeling of the object
being reconstructed, followed by an entirely separated rendering step.
We want to allow for an interactive model--render--compare feedback loop in
order to speed up and simplify the modeling process.

These problems lead us to the specification of a mesh-based
interactive construction software similar to commercial modeling tools.
The main difference between our tool and conventional 3D animation packages
is that \emph{Shape} produces not only images
but also spectral information which is presented in a way to be directly comparable
to various types of astronomical observations,
and that its modeling environment focuses on the structures and symmetries that are
commonly present in astronomical objects.

The \emph{Shape} program consists of two main components:
A 3D modeling view in which geometry and behavior of the model are defined,
and a 2D screen where the simulated appearance of the model can be compared to observational data (Figure~\ref{shape_gui}).

The model properties are defined basically in three steps.
First a structural selection of regions in the spatial domain is set using mesh surfaces and volumes.
Multiple meshes can be used to model complex structures.
Physical properties are assigned as a function of position in the spatial domain.
Different functions can be used for different sub-structures, i.e. meshes.
A sampling process in the selected spatial mesh regions then reads the physical properties.
The sampling can either be done at particle locations on the surface or within the volume of the
mesh or at the locations of the voxels of a regular 3D grid.
Sampling particles can be distributed randomly on the surface or the volume of a mesh.
The particles can also be used to visualize the vector field, attaching to them a velocity vector.
Although these particles have physical attributes, they should not be confused with physical atomic
particles.
Here they serve only the purpose of visualization and of defining a position for the sampling process.

For the mesh creation process, a number of suitable primitives (sphere, torus, cone, cube, etc.)
are available which are then deformed using a selection of \emph{modifiers};
importing meshes from other software is also possible.
The primitives can be used either as volumetric objects, as infinitely thin shells,
or as volumetric shells with user-specified thickness.
In addition to scaling, translation and rotation,
the available modifiers include operations that are parameterized along a given axis
(the squeeze, squish, shear and twist modifiers).
Boolean combination of different primitives allows for constructive solid geometry modeling.
Additionally, mesh vertices can be adjusted manually either individually or in groups.

Among others, the spatial distribution of density, color and velocity
can be conveniently defined as a function of position.
The user can choose among a set of predefined common spatial dependencies with adjustable parameters
or create custom functions in different coordinate systems
(cartesian, cylindrical or spherical).
These functions are assumed to be separable into functions of the chosen coordinates,
e.g. $f(r, \theta, \phi) = f_r(r) \cdot f_\theta(\theta) \cdot f_\phi(\phi)$.

The different coordinate functions ($f_r$, $f_\theta$ and $f_\phi$, in that example) can be defined either
analytically or as piecewise linear functions that may be graphically edited
(like animation curves in conventional animation software packages).
For the analytic functions, the user can define the function by typing in a formula.
The formula is fed into a mathematical text interpreter and is calculated and displayed
immediately as a graph.
Predefined functions for, e.g., the velocity field may include common movements such as
radial expansion, gaseous disk rotation and solid body rotation, and random distributions.
More complex functions can be constructed by sequentially combining them
with addition or multiplication.

All modeling and visualization can be done in either arbitrary units or in various
actual physical units that are suited to the problem,
e.g. the apparent size of the object in arcseconds, sizes and distances in parsec or
astronomical units, velocities in km/s and the like.
The modeling interface not only displays the mesh and particles, but also
the particle velocities in the form of vectors color-coding their velocity
along the line of sight, which is responsible for the observed
Doppler-shift.
The projection of these
vectors on the sky is directly observable in some objects as tangential motion,
after comparing observations with a sufficiently large time interval in between.
This provides further kinematical constraints
(see also Section \ref{saturn}, Figure~\ref{ngc7009_vectors}).

Since the particle velocities can be specified in a model,
the time evolution of the object may be predicted,
assuming ballistic expansion (i.e. constant velocities).
A time modifier calculates the future or past positions of the particles
for a given time interval.
Using this feature, the age of a nebula and its short--term structural evolution can be estimated.

Finally, our tool also allows importing data from external simulations in order to
visualize and analyze the results (cf. the examples section).
This makes it also possible to use other software for specialized modeling tasks
(e.g. for the generation of sophisticated noise distributions).
Also, the models created with \emph{Shape} may be exported and used as input
for external simulation or visualization software.

\subsection{Image Rendering}

Several renderers are available, which serve different purposes according
to the type of object that is modeled and the adopted workflow.
Many astrophysical objects are optically thin, i.e. transparent.
This fact has been used in three renderers (\emph{particle},
\emph{grid} \& \emph{mesh} renderers). They provide faster rendering than the \emph{physical} renderer,
which takes into account opacity and other radiation transport effects.
The \emph{physical} renderer is still experimental and under development.
Its details will be described elsewhere.

The \emph{particle} renderer uses a random particle distribution to sample
the model emissivity and velocity space.
The values are added directly to the image pixels or P-V diagrams according
to their projected position and velocity along the line of sight.

In the \emph{grid} and \emph{mesh} renderers, a regular grid is set up in world space
that is aligned with the line of sight.
In the \emph{grid} renderer, particle positions are used to sample
the physical properties of the object.
The density of the particles are distributed in the 3D grid.
The voxels are then rendered and/or output for external purposes.
For the \emph{mesh} renderer, the sampling of the physical properties is
done by searching for the voxels of the grid that overlap with the model`s mesh.
If the mesh is a volume, then the center point of the voxel is
used as the sample point and
emission is calculated from the fraction of the voxel that is located within the mesh.
To reduce aliasing effects, the position can be jittered within the limits of the voxel.
If the mesh is a surface, the fraction of each mesh segment in a
particular voxel is determined.
Assuming a small but finite thickness of the surface,
the fraction of the volume occupied in the voxel, and hence
the emission, can be determined.
The emission from each sub-object is added to the grid and rendered separately
(no mixing is done).
Finally, the emission from each voxel is projected onto the image plane.

In the images, the emission is integrated along the line of sight ($z$-axis)
regardless of
their velocity, and the results are plotted in the image plane ($xy$-plane).
Position-velocity diagrams (P-V diagrams), however, only take emission
in a given $x$ range (within the spectrograph slit) into account.
The intrinsic spectral line width is assumed small compared to the resolution
of the P-V diagram.
The $y$ coordinate is the position along the slit.
The object's emission is
distributed according to its position along the slit and the velocity component along the line
of sight ($v_z$, see Figures \ref{doppler} and \ref{image_pv}).
\begin{figure}[t]
\centering
\includegraphics[width=0.9\columnwidth]{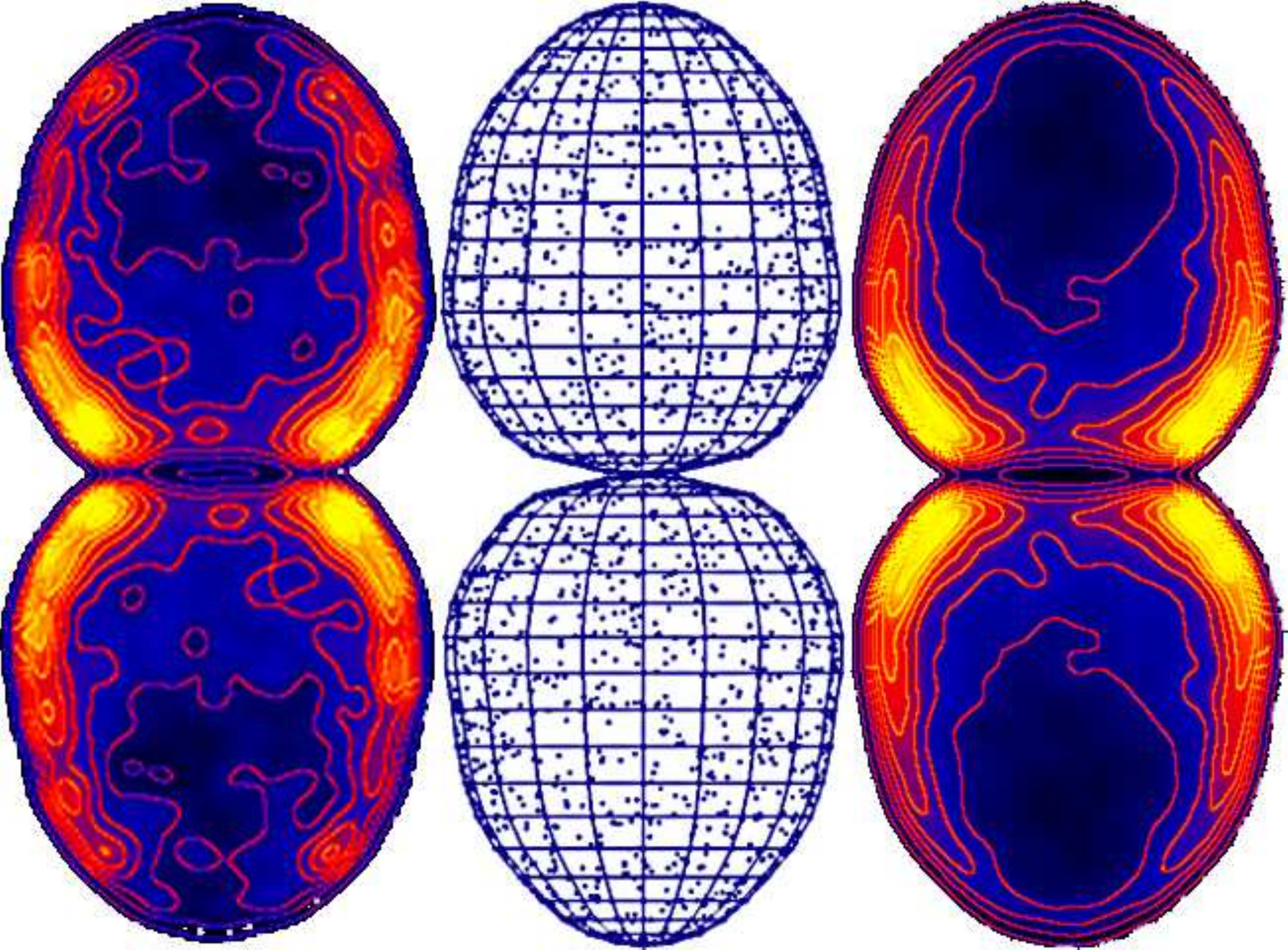}
\caption{
The image on the left is a rendering of a hydrodynamical simulation after
being imported into \emph{Shape}.
The middle shows the \emph{Shape} mesh fitted to this image and
a small fraction of the sampling particles.
The panel on the right is the rendering of the \emph{Shape} model of the hydrodynamical
simulation
after manually fitting an emissivity distribution in cylindrical coordinates~\cite{WSGGS2009}.}
\label{hydro_mesh}
\end{figure}

\begin{figure}
\centering
\includegraphics[width=\columnwidth]{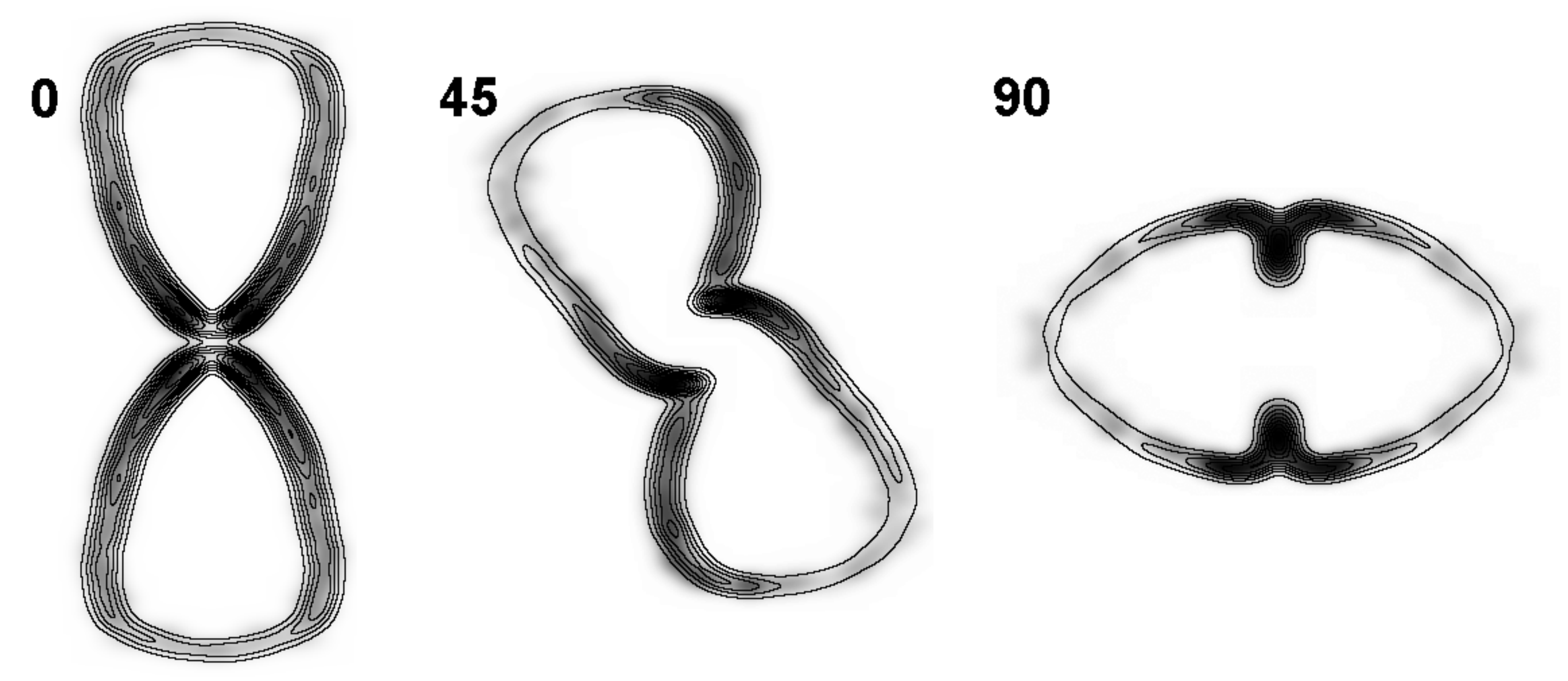}
\caption{Position--velocity diagrams are shown for the hydrodynamical simulation
(grey scale) and \emph{Shape} model (contours) of Figure~(\ref{hydro_mesh})
for three different viewing angles.
Note that the outline changes with the viewing angle.
Furthermore, the outline of the
first P--V diagram, which corresponds to the viewing angle used in the rendering
of Figure~(\ref{hydro_mesh}), is different from that of the image.
The middle P--V diagram shows point-symmetry, whereas the others are mirror-symmetric.
Both observations are indications for deviations from a homologous expansion of the nebula~\cite{WSGGS2009}.
This shows that reconstruction methods that rely on homologous expansion will yield
qualitatively inaccurate results for such a nebula.}
\label{hydro_pv}
\end{figure}

\subsection{Synthetic observations}

An important task in the process of reconstructing an object based on observational data
is, of course, the comparison between these data and the model.
In order to make this comparison reliable, the rendering algorithm has to take
into account not only the model itself, but also the properties of the measurement devices.

For comparison with observations, {\em Shape} includes three main 2D data types
that are rendered from the 3D model: images, P--V diagrams and channel maps.
For the images and P--V diagrams, several coloring schemes are available
that display different types of secondary information.
First, the color of a substructure can be used as assigned in the
3D model. This helps to distinguish the different sub-systems of the object,
especially in the spectra.
Second, the velocity along the line of sight as given by the Doppler-shift
can be color-coded as simple red/blue to distinguish regions
that move towards or away from the observer, respectively.
Another option maps the Doppler-shift
to a rainbow color range, such that a more detailed appreciation of
the line-of-sight motion can be achieved.

Figure (\ref{image_pv}) shows the rainbow display option for image (a)
and P-V diagrams (b \& c).
In panel (a) the image of a synthetic bipolar structure is shown as seen with a
structure and inclination angle similar to the outline in the schematic
diagram of Figure (\ref{doppler}).
The color-coding maps the Doppler-shift to the rainbow colors from red to violet,
with red receding and blue approaching the observer.
The width of the spectrograph slit has been marked.
Panel (b) of the same figure is the resulting P--V diagram with a scaling in velocity (horizontal)
that is typical for high-resolution spectral observations of bipolar planetary nebulae.
In this object the velocity vector is proportional to the position vector (homologous expansion),
such that for a suitable velocity scaling, the P--V diagram should reproduce the object structure
along the line of sight.
This is confirmed in panel (c) where the velocity scale has been stretched accordingly.
If there are deviations from a homologous expansion, the structure of the P--V diagram can not
be stretched to yield an accurate representation of the object structure along the line of sight
(see the examples section below).

A third image type that is common in astronomy is the so-called \emph{channel map}
where only a given velocity (or wavelength) range is projected onto the $xy$-plane;
\emph{Shape} is able to produce this output, optionally presenting maps
for many different velocity ranges next to each other.
This observing mode is typical for radio wavelength observations of spectral lines of molecules
in dense star forming regions and young planetary nebulae.

The channel map mode in \emph{Shape} can also be used to display light echoes,
where each image is a channel of equal light travel time.
Such light echoes may occur when a star experiences a sudden and short outburst.
The light of the outburst may then be reflected by surrounding dust.
The appearance of the nebula changes on a time scale corresponding
to the time it takes to propagate from the star through the surrounding dust.
At a given time, we on Earth can only observe the radiation that has taken the
same amount of time from the star to the nebula and then to Earth.
Therefore, the light echoes represent channels of equal light travel time.
The geometrical shape of such a light echo in space is that of a paraboloid.
It has the property of equal distances from the position of
the star (the focus) to a given position on the paraboloid and then
to the observer.
The channel map mode allows one to calculate and display such light echo images.
A well known example of such a light echo is the one observed around the star V838 Monocerotis
after its large outburst in 2002~\cite{Bond2003}.

Some instrumentation only registers particular spectral ranges or has a varying
sensitivity at different wavelengths.
Therefore a set of filters can be applied during the rendering process,
that filters the emission
according to a customizable set of rendering filters.
These adjust or exclude emission according to the specifications of the filter
(see below, the example of nova RS Ophiuchi).

Final post-processing steps on the rendered raw image and spectral information are
used to model the properties of observational devices,
e.g. the influence of the atmosphere or the aperture.
The most important of the atmospheric and instrumental effects are the spatial and
spectral resolution.
These are modeled by applying gaussian convolution to the spatial and spectral dimensions.
Effects like complex point-spread functions and ghost images from reflections inside
the instrumentation are not taken into account.
Other filters like gamma correction or logarithmic scaling make high dynamic range
data manageable.
Contour levels of the image intensity may also be applied.
The rendered image and spectrum can then be displayed and interactively blended
with the observation to visually detect differences.

When the model satisfies the demands of the user,
a number of visualizations can be done that focus on the presentation of the results
instead of being a modeling aid.
For example, images may be rendered from different viewpoints, and
image sequences and videos may be generated by changing display parameters
(e.g. the camera position or the slit position of a P--V diagram).

\subsection{Automatic Optimization}

\emph{Shape} includes an automatic optimization module
which minimizes the difference between the rendered model and the observational data
by tweaking the values of observing parameters or those of the modifiers in the model.
The user can influence this process by selecting the parameters to be optimized and
by limiting their values to a plausible range,
as well as by optimizing different parameters in order of priority until
the model adequately fits the observations.

Since a model may contain hundreds of parameters that influence each other,
the user is presented with a hierarchical list of optimizable parameters
from which a reasonable subset (usually only a few parameters) are then selected for optimization.
This interactive parameter pre-selection not only speeds up the following algorithm,
but it may also be seen as an additional way for the user to emphasize certain properties of the model:
Since changing the value of any one of two independent parameters might make the model
explain the observational data equally well,
there is ambiguity in any optimization process that can only be resolved by the user.
Specifying ranges of valid parameter values is an even more powerful method of user intervention,
because not selecting a parameter for optimization is eventually equivalent to selecting
an infinitely small value range.
Many parameter types presuppose certain boundaries for their values.
For example, in a squeeze modifier which scales the distance of vertices from a given
axis, the scaling coefficients are necessarily positive.
For the optimization one can use images, P--V diagrams or data plots,
of which only one can currently be optimized at a time.

\subsection{Automatic reconstruction}

Although \emph{Shape} has been conceived for interactive reconstruction, there is an automatic
reconstruction module.
It is intended to provide a first look at the structure of a complex object
for which a suitable data set is available.
From the input data it generates a particle system that follows the brightness distribution of the data.

Our system has several advantages compared to other automatic reconstruction systems that are based
on a mapping between Doppler--shift and position along the line of sight.
First, it can use two different data types: P--V diagrams and channel maps.
Second, it is not restricted to the assumption of a homologous expansion.
It can use any radial velocity law as long as it is monotonic and provides
a one-to-one mapping from Doppler--shift to position along the line of sight.

The input data in the form of P--V diagrams, in contrast to the tomographic method by
Sabbadin et al.~\cite{Sabbadin06:SPN}, are assumed to come from parallel slits.
Parallel slit settings are much easier to obtain and process,
since they can cover the complete object without gaps or superpositions.
Similarly, a set of channel maps can be used to recover the 3D structure.
Currently, no interpolation is performed if there are gaps in the
spectrograph slit coverage.
The gaps can, however, be filled by assuming that the slits are wider
and touch each other.
Slit positions and widths for the input P--V diagrams can be set individually.
As usual, the exact scaling between Doppler--shift and position must come from
some symmetry information about the object or some part of it.
It is clear that such a reconstruction can only provide a first approximation to a detailed
interactive reconstruction,
especially if the assumed velocity law is not accurate for any part of the object.

\subsection{Plot, Animation and Movie Modules}

There are three auxiliary modules in \emph{Shape} that are fully integrated in the main modeling process.
They are the modules for plotting data, animating model parameters and the display of animation
results.

The plot module can display graphs of a variety of data from the models, including 1D spectral line
profiles.
While the spectral line profiles require the rendering of the model,
other data types can be obtained directly from the particles in a 3D model.
For instance, when plotting velocity along the line of sight as a function of
position along the spectrograph slit,
a real-time preview of the P--V diagram is obtained in the form of
a scatter plot of these particle properties.
Such data can be visualized in real-time, as the user changes, for instance,
the camera orientation or the slit position in the other modules.

The animation module was inspired by the equivalent curve editor modules of
conventional 3D animation software.
Here almost all object and camera parameters can be animated as a function of
time using either manual animation curve editing or algebraic functions.
In addition to educational visualizations, this allows the scientist to
effectively explore the
parameter space by producing sequences of outputs by automatically changing
the parameters in a controlled manner.
The animation module has an interactive time-line that
updates the object structure in the 3D module in real-time.
After rendering a sequence, the results can be viewed in the movie module.
The movie module can load and simultaneously reproduce an arbitrary number of animation sequences,
which may include images, spectra and plots.
This is very helpful when analyzing various types of outputs as the parameters of a model change.

\section{Results and Example Applications}
\label{results}

In this section we show three examples of \emph{Shape} models that have been
published in the astrophysical research literature.
The first example is an analysis of hydrodynamical simulations.
These have been used to validate \emph{Shape} and extract information about
deviations from a homologous expansion that can be expected in planetary nebulae.
Using \emph{Shape} these deviations have been reduced to a small set
of parameters for later use in models of real objects, like the one of the
planetary nebula NGC~7009, which is presented after the hydrodynamical simulation.
The third example is that of Nova RS Ophiuchi, which has been constructed by
a user that is independent of the \emph{Shape} developers.
It applies features that have been implemented upon request by the user.

For this work, \emph{Shape} has served as a completely integrated tool for
modeling, simulation, analysis and visualization, setting it apart from any
existing commercial or research tool.
The results shown in this section represent original astrophysical research
results which would have been very hard or impossible to obtain without
\emph{Shape}.

\subsection{Validation of Shape with Hydrodynamical Simulations}

In addition to direct modeling of astrophysical observations, \emph{Shape}
has been applied to the analysis and visualization
of morpho-kinematical aspects of numerical hydrodynamical simulations.
Such simulations have also served to validate the functionality of \emph{Shape}, since
their properties are known in full.
Steffen \& Garc\'{\i}a-Segura~\cite{WSGGS2009}
have used \emph{Shape} to characterize the velocity field of numerical simulations
of some basic types of planetary nebulae.

Parametric descriptions of the velocity field that extend
commonly assumed velocity fields in morpho-kinematical modeling of axisymmetric objects
where derived.

For this analysis, the simulated hydrodynamical data have been filtered according to density.
The densest elements correspond to the shell that is usually observed as the brightest
region in a planetary nebula.
They have been imported to \emph{Shape} as a particle system,
including their velocity information.
A 3D mesh and density distribution was then fitted to the large-scale structure.
Figure (\ref{hydro_mesh}) shows a comparison of the integrated emission measure (i.e.
density squared) from one of the simulations (left) and that of the corresponding
\emph{Shape} model (right).
The mesh structure and particle distribution (10 \% of the actually used particles)
are shown in the middle panel of the same figure.

The authors allowed the presence of a poloidal velocity component.
This component is perpendicular to the radial velocity component and points towards
the symmetry axis along the ``longitude'' of a spherical coordinate system.
As a function of angle from the symmetry axis, the poloidal velocity can be described
by three linear segments with the condition of zero magnitude on the symmetry axis and equator.
The radial component can also be described by two or three linear segments.
These descriptions of velocity fields capture the kinematical properties of nebulae quite accurately,
and are simple enough to be easily implemented in morpho-kinematical or
photoionization codes (Figure~\ref{hydro_pv}).

To verify the kinematic modeling functionality of \emph{Shape},
the velocity field of the simulations was manually fitted to high accuracy
including small-scale variations.
For later use in models of actual observations (see subsection \ref{saturn}),
the radial and poloidal velocity field was simplified to two or three linear sections
depending on the particular model.
In Figure~(\ref{hydro_pv}) in grey--scale we show, for three different viewing angles,
the P--V diagrams of the simulation that was also used in Figure (\ref{hydro_mesh}).
Note the difference in the shape of the first P--V diagram from that of the image
renderings in Figure~(\ref{hydro_mesh}, same viewing angle).
This is a clear indication for deviations from a homologous expansion.
For a homologous expansion, a change in viewing angle should show only a rotating
structure even in the P--V diagrams.
The changing structure at different viewing angles confirms the presence of deviations.
In a reconstruction that incorrectly assumes a homologous expansion,
the derived 3D structure will depend on the viewing angle with deformations
along the line of sight.

The contour lines in Figure~(\ref{hydro_pv}) are those of the
simplified velocity field model.
Together with the reconstructed structure in Figure~(\ref{hydro_mesh})
they show that piecewise linear velocity fields can yield an accurate representation
of the global velocity structure in commonly seen planetary nebulae.
In the following section we show how this result has been applied to a planetary nebula.

\subsection{Reconstruction of the Saturn Nebula}
\label{saturn}

\begin{figure}[t]
\centering
\includegraphics[width=\columnwidth]{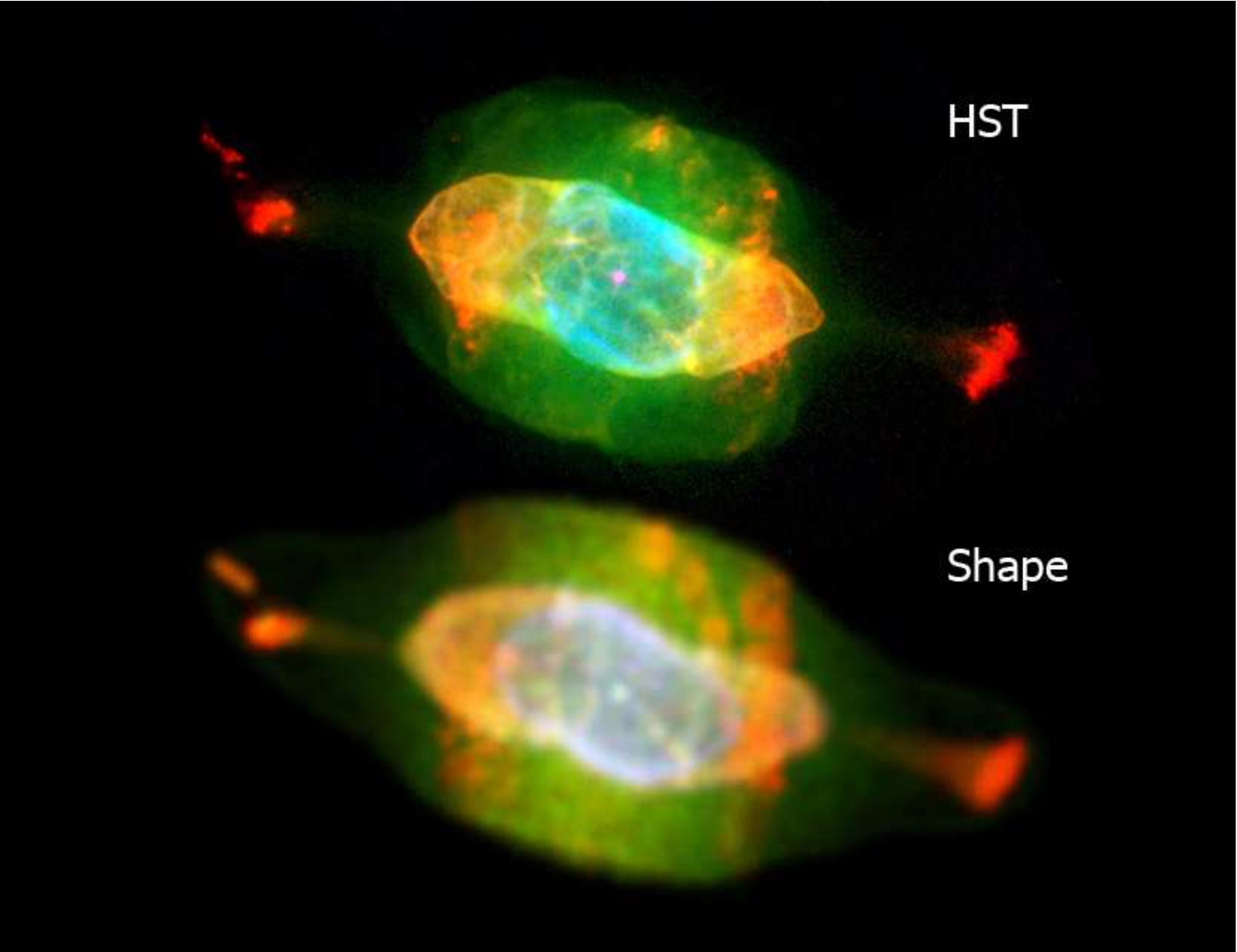}
\caption{A research example: The planetary nebula NGC~7009
as observed by the \emph{Hubble Space Telescope} (top)
and as modeled with \emph{Shape} by an astronomer (bottom).
Different colors
represent different spectral line images. They have been modeled
with similar structural meshes, but quite
different emissivity distributions.
This model is based on the
one published in~\cite{Steffen09}.}
\label{ngc7009_hst_shape}
\end{figure}

We exemplify the reconstruction of complex planetary nebulae with the
case of NGC~7009~\cite{Steffen09}, also called the \emph{Saturn} nebula.
The top panel of Figure~(\ref{ngc7009_hst_shape}) shows a color image
of the nebula as constructed from three narrow-band filter images
obtained with the Hubble Space Telescope (HST)~\cite{Balick98}.
In the bottom panel a \emph{Shape} model of the nebula is shown.
This is a version of the model that includes more structural details than the one
published in by Steffen et al.~\cite{Steffen09}.
Compared to the model in that publication, for the rendering in Figure~(\ref{ngc7009_hst_shape})
we have added more small-scale features and models of additional narrow-band lines.

To model the object we took into account ground-based spectral data of four different
spectral lines~\cite{Sabbadin04} in addition to color HST images.
The spectral data are composed of 12 position-velocity diagrams rotated at intervals
of $30^\circ$ around the central star
with a spatial resolution of approx. 1~arcsec and a velocity resolution of about 7~km/s.

Sabbadin et al.~\cite{Sabbadin04} used these data to derive the structure of NGC~7009 with their
tomographic method which assumes homologous expansion (see section \ref{automatic}).
Since the spatial coverage of the spectral data is incomplete,
they apply an angular interpolation in regions of missing data between slits.
This produces a unique solution for the 3D structure, but at the spatial resolution corresponding
to ground-based observations, which is more than a factor of 5 worse than that of the HST.
Although the object is not exactly axisymmetric,
the reconstructed 3D structure of the object shows deviations from
axisymmetry that are similar to those introduced by deviations
from a homologous expansion.

Furthermore, the presence of hot X-ray emitting gas within the inner
shell~\cite{Guerrero02} leads to the expectation that this shell
is likely to present deviations from a non-homologous expansion.
The deviations have been estimated from a comparison between P--V diagrams and images.
Although no unique solution was found, the distortions of the structure introduced by assuming a homologous
expansion have been reduced considerably by the proposed velocity fields which
include a poloidal velocity component~\cite{WSGGS2009}.
They are within the expectations from the earlier hydrodynamical simulations.

\begin{figure}[t]
\centering
\includegraphics[width=\columnwidth]{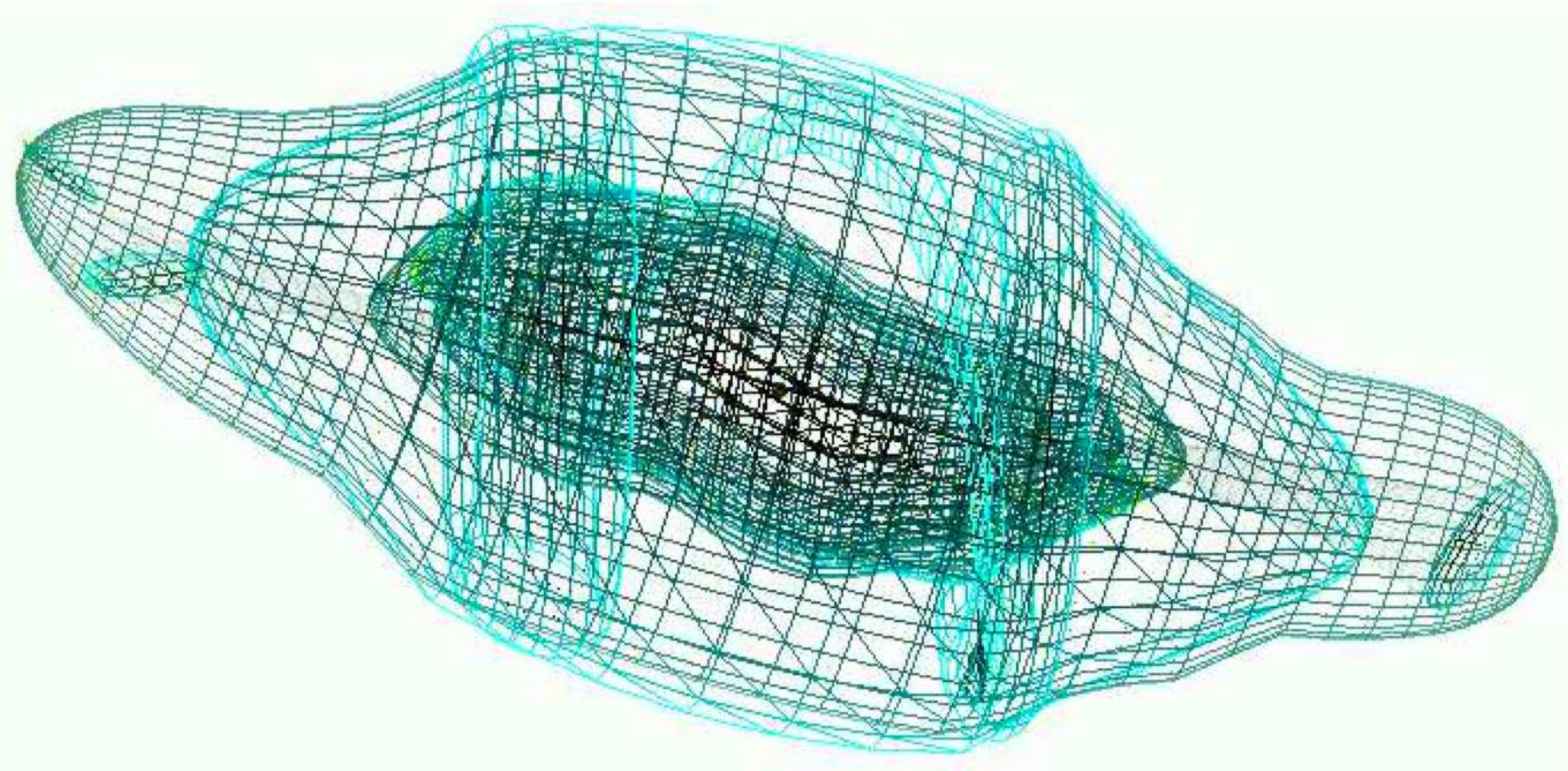}
\caption{In \emph{Shape}, models are constructed from \emph{meshes}
which are generated from simple primitives
and deformed using various modifiers.
This model of the \emph{Saturn nebula} (NGC~7009)
has a number of nested transparent shells~\cite{Steffen09}.}
\label{ngc7009_mesh}
\end{figure}

In contrast to the direct tomographic reconstruction, we interactively modeled the structure
as a set of nested meshes (Figure~\ref{ngc7009_mesh})
and compared the rendered images and P--V diagrams with the observed counterparts.
The interactive flexibility in the modeling of the complex structure and velocity field
allows it to readily take into account complex structures without the need of any new coding.
Volume meshes have been used for the large--scale structures, whereas
the thin main shell was modeled as a surface mesh.
Knots and filaments were reproduced using manually added particles on a copy of the
main shell.
The velocity and emissivity distributions have been adjusted separately for each mesh
using piecewise linear functions.

\begin{figure}[t]
\centering
\includegraphics[width=\columnwidth]{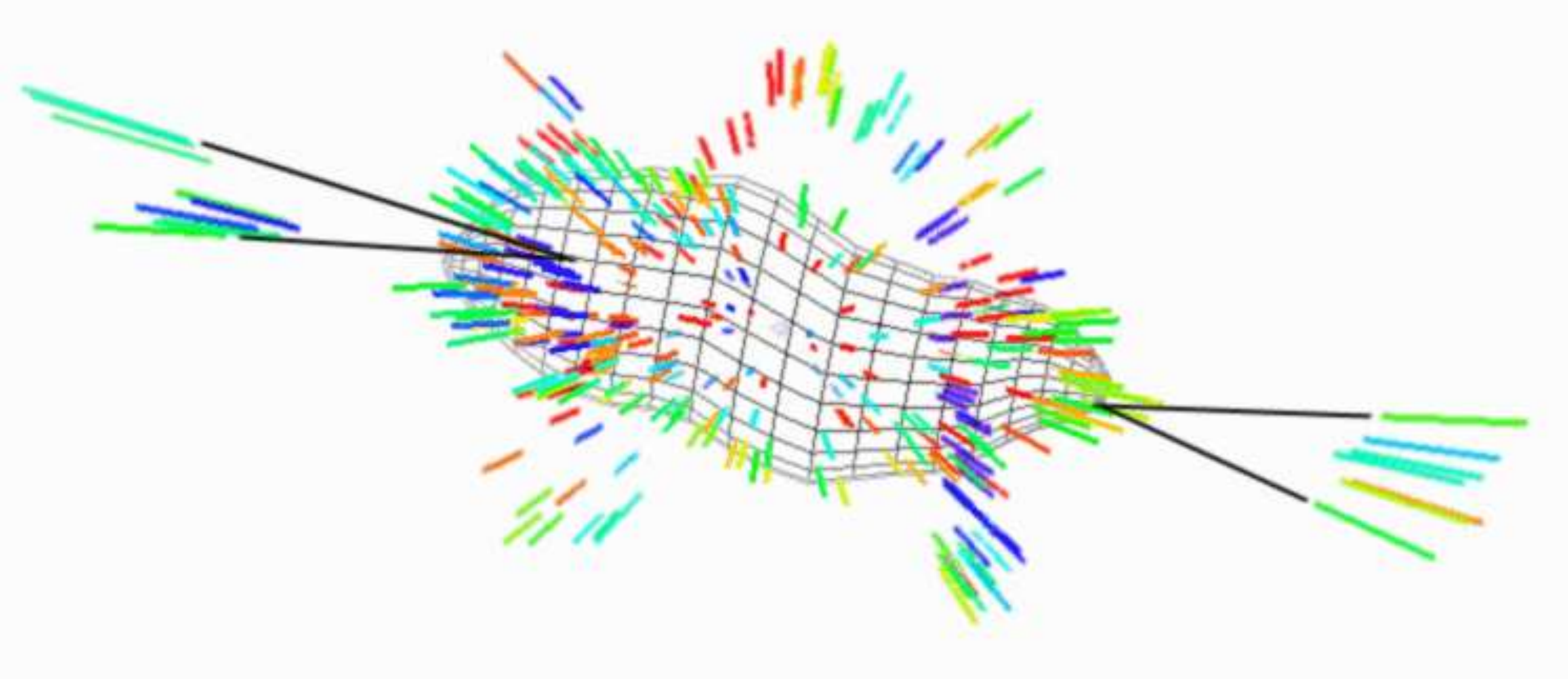}
\caption{The velocity vector components in the plane of the sky are shown for
the model of NGC~7009~\cite{Steffen09}.
The color encodes the velocity component along the line
of sight, with blue approaching and red receding from the observer.
While the expansion along the line of sight is observable directly
using the Doppler-shift on spectroscopic emission lines, observing
the velocity component in the plane of the sky requires the detection of
position shifts in images that have been obtained with a considerable
time difference (usually more than five years with the HST).
The reconstruction of the velocity field as shown is a direct prediction for the
expected expansion pattern of NGC~7009.
After its observation it can provide evidence for or against the model.}
\label{ngc7009_vectors}
\end{figure}

Observationally, tests of the model can be achieved by measuring the expansion component
in the plane of the sky (which can not be obtained by spectroscopy).
Unfortunately, this requires the detection of the expansion in images that have been obtained with a time difference of a decade or more with the \emph{Hubble Space Telescope}.
For NGC~7009 this has not yet been adequately achieved.
The velocity vector visualization in \emph{Shape} provides an immediate built-in
prediction of the expansion pattern in the plane of the sky.
Figure~(\ref{ngc7009_vectors}) shows the projected velocity vectors for one of the models
of NGC~7009.
They clearly show the deviations from a homologous expansion, in that the vectors of
the main shell and the bright symmetric knots do not converge at the position of the
central star.
Future observations of this pattern will provide hard evidence in favor or against
the model published in Steffen et al.~\cite{Steffen09}.

\subsection{Nova RS Ophiuchi}

\begin{figure}[t]
\centering
\includegraphics[width=\columnwidth]{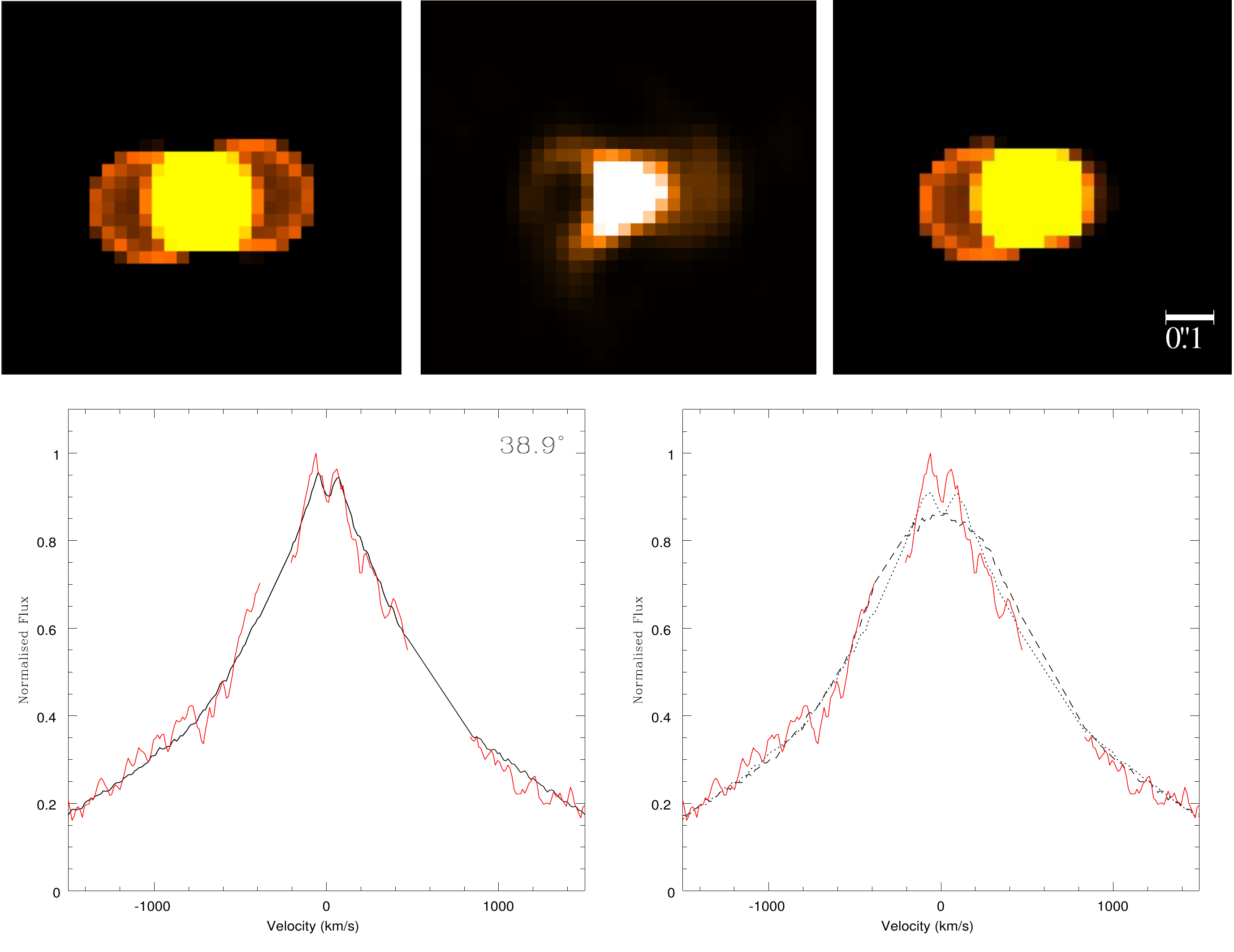}
\caption{The top panels show the observed image (center) and two different
model renderings of nova RS Ophiuchi~\cite{Ribeiro2009}.
The model
on the right includes a special rendering filter that corresponds to the
transmission filter on the \emph{Hubble Space Telescope}.
It excludes
some of the emission with very high velocities along the line of sight.
The bottom panels show the spectral line profiles from the observations
(noisy red line)
and the final model (left, continuous line).
The right panel compares the
observed profile with models at different viewing angles that are just
consistent with the observations (dashed and dotted lines).
[Figure reproduced with permission of the authors (Ribeiro et~al.~\cite{Ribeiro2009}).]}
\label{ribeiro_7}
\end{figure}

Ribeiro et al.~\cite{Ribeiro2009} have used \emph{Shape} to model the structure
and expansion of nova RS Ophiuchi after its outburst in 2006.
In this system a white dwarf star orbits inside the outer layers or stellar wind
of a red giant star.
The white dwarf accumulates material from the giant, which after some time
produces a nuclear explosion on its surface.
The result is a fast expanding shell around the binary system (Figure~\ref{ribeiro_7}, top).
Nova explosions may have speeds of several thousand kilometers per second,
which produces considerable Doppler--shifts in the observed spectral lines
(Figure~\ref{ribeiro_7}, bottom).
Some emission might then be outside the range of observations with narrow-band filters on
the \emph{Hubble Space Telescope} and may therefore go undetected.
Ribeiro et al.~\cite{Ribeiro2009} have used the spectral rendering filter in \emph{Shape}
to explain such "missing'' regions in their \emph{HST} imaging observations of Nova RS Ophiuchi.
This feature was introduced in the software upon their request.
When they do not use the HST filter transmission, the object is symmetrically double-lobed
(Figure~\ref{ribeiro_7}, top, left).
When the filter transmission is included, one of the lobes largely disappears
(Figure~\ref{ribeiro_7}, top, right).
The model with the filter matches the observed image much better, although there is still
some discrepancy in the detailed structure of the larger lobe (Figure~\ref{ribeiro_7}, top, center).
The inner bright region dominates the 1D line profiles (Figure~\ref{ribeiro_7}, bottom) and was
very useful to set limits on the orientation of the nebula.

\begin{figure}[t]
\centering
\includegraphics[width=\columnwidth]{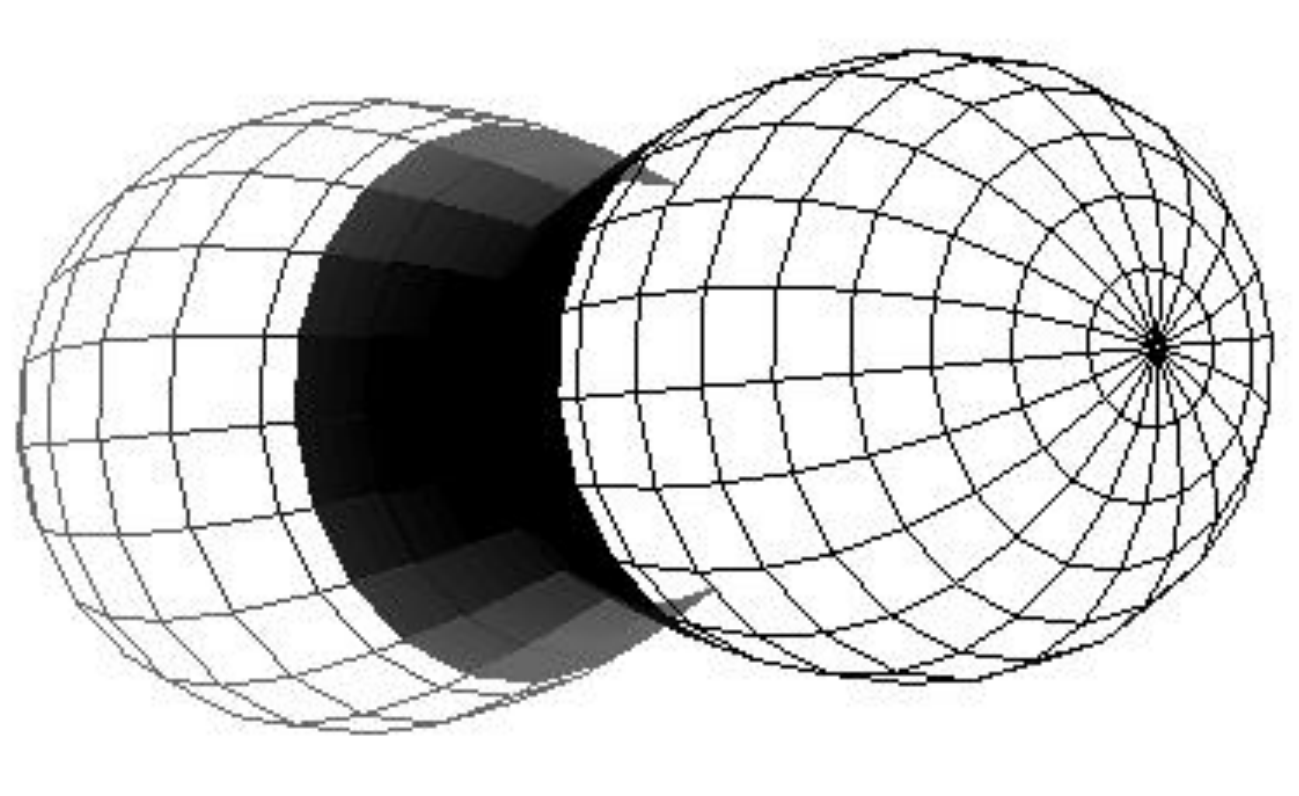}
\caption{The outburst of nova RS Ophiuchi has been modeled by Ribeiro et~al.~\cite{Ribeiro2009}
using an expanding bipolar nebula (mesh structure) with a nearly inert waist
(shaded structure).
Using the time modifier feature of \emph{Shape} they
were able to model the expansion between the first and second epoch observations.
[Figure reproduced with permission of the authors (Ribeiro et~al.~\cite{Ribeiro2009}).]}
\label{ribeiro_6}
\end{figure}

Figure~(\ref{ribeiro_6}) shows the mesh of the bipolar nebula that they constructed.
The inner shaded region was found not to expand significantly during the explosion,
as shown by second epoch observations.
The top panels of Figure~(\ref{ribeiro_6}) show the rendered images on
the right and left that do and do not include the spectral rendering filter, respectively.
They can be compared with the observed image in the middle.
The bottom panels compare the synthetic spectral line profiles of the final
model (left) with the observed line profile (noisy line).
On the right, the range of model line profiles is shown that is still compatible with
the observations.
The difference between them is only due to the viewing angle in the range from $29^\circ$ to
$40^\circ$ deviation of the object axis from the plane of the sky.

Using the time modifier with the assumption of
ballistic expansion, Ribeiro et al.~\cite{Ribeiro2009} have been able to show that there is a
considerable difference in the expansion of the bright inner region and the dimmer
bipolar lobes.
In this developer independent work, the multi-functional interactive modeling approach
has proved to be especially fruitful.

\subsection{Content production for digital media}

Since their emergence, commercial animation systems have been applied to produce
animated visualizations of high spatial and temporal resolution of astronomical phenomena
in scientific documentaries and feature films.
More recently, astrophysical research simulations have contributed impressive
visualizations of phenomena that are impossible to do in commercial animation packages.
A stunning example is the evolution of a star forming region simulated by Henney et al.
that was presented as part of the show ``Journey to the Stars'' by the Hayden Planetarium in New York
and other digital planetaria.

\emph{Shape} has the potential to simulate and visualize a variety of astrophysical phenomena
for graphical media applications, both in animation and stills.
A few examples of such visualizations can be seen on the \emph{Showcase} page of
the \emph{Shape} website (\url{http://www.astrosen.unam.mx/shape}).
With the future development of our software and that of computing resources, \emph{Shape} will
become a very useful tool to visualize and illustrate astrophysical phenomena for a
variety of media, including print, television and digital planetaria.

\section{Future Development}
\label{future}

The current and near future development of \emph{Shape} is steered by two driving forces:
The first is the type of scientific applications that the developers and current users are
working on.
Second, development constantly addresses the existing limitations of the software.
Limitations include the availability of processing memory in
the \emph{Java Virtual Machine} which mainly translates into limited spatial and spectral
resolution.
Another significant limitation is that texture mapping is currently not available, which
would boost the possibilities to model noise structures like complex filaments that
are observed in many nebular objects.
Since the development of the \emph{Java} version of \emph{Shape}
began, new features and changes have been introduced continuously.
Current developments include explicit radiation transfer for dust scattering
and spectral lines.

Long-term plans are guided by \emph{potential} applications that the system has.
Such plans include the incorporation of interactive hydrodynamical simulations
making use of multi-processor graphics processing units.
Gravitational interaction for particle systems in \emph{Shape} will
allow it to simulate a variety of phenomena, like interacting galaxies and multiple stars.
A substantial increase in spatial resolution will make the system applicable to
very realistic modeling and rendering of astrophysical objects for
educational purposes in planetaria and other electronic media.

\section{Conclusion}
\label{conclusion}

We have presented a novel 3D application for the modeling and reconstruction of
astrophysical objects that incorporates interactive modeling tools.
It considerably extends the capabilities of conventional reconstruction systems.
This is achieved through the support of a system of construction and
``modifier'' tools that allow
extremely complex structures and velocity fields to be assembled without the
need for user programming.
A number of visualization styles that are common
for astronomical observations can be used to compare the model with the observed data.
The workflow is enhanced by the ability to continuously compare the model
to observational data during the modeling process as well as by an automatized
optimization algorithm.
The tool has been shown to cover the entire modeling and visualization pipeline of a
common morpho-kinematical modeling task,
supporting the scientifically accurate reconstruction of a wide class of astrophysical
objects while keeping a convenient and user-friendly interface.
The software we have presented has been thoroughly tested and applied in a number of
astronomical research projects, some of which we have quoted as examples.
In addition to scientific research, it may prospectively be applied for physically
plausible artistic works or for the generation of astronomical animations for educational
purposes, e.g. in digital planetariums.
Non-astrophysical uses can also be imagined wherever velocity information is observed,
e.g. in the field of Doppler radar observations of tornados and other weather phenomena.

\emph{Shape} is freely available as a Java WebStart application from its website at \url{http://www.astrosen.unam.mx/shape/}

\newpage

\section*{Acknowledgments}

The authors would like to thank the reviewers for their insightful comments and constructive criticism.
W.~S. and N.~K. acknowledge support from CONACYT grants 49447 and UNAM DGAPA-PAPIIT IN108506-2.
W.~S., C.~M., S.~W. and M.~M. have been supported by DFG grant 444 MEX-113/25/0-1. 
S.~W. and M.~M. acknowledge support by DFG grant MA 2555/7-1. 
N.~K. acknowledges support from the Natural Sciences, Alberta Ingenuity Fund and Engineering Research Council of Canada and from the Killam Trusts.


\bibliographystyle{IEEEtranS}
\bibliography{IEEEabrv,shape}


\begin{IEEEbiography}[{\includegraphics[width=1in,height=1.25in,clip,keepaspectratio]{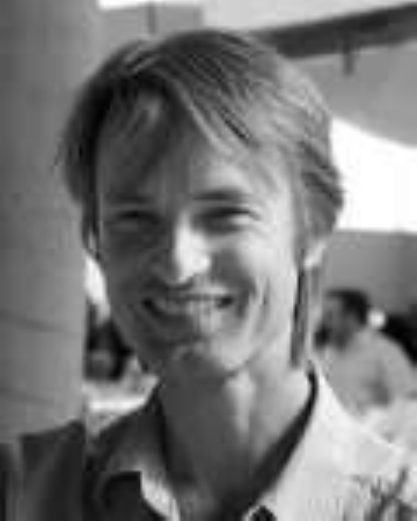}}]{Wolfgang Steffen} is Professor of Astrophysics at the Institute of Astronomy of the Universidad Nacional Aut\'onoma de M\'exico in Ensenada, Mexico. He received his BA~(1992) in Physics and the PhD~(1994) in Astrophysics, both
from the University of Bonn and the Max-Planck-Institute for Radioastronomy in Bonn. He did his postdoctoral work at the University of Manchester (Department of Astronomy) where he developed an early version of \emph{Shape}. From 1998 to 2002 he was at the University of Guadalajara (Mexico) where he began \emph{Proyect Cosmovisi\'on} to develop realistic
computer graphics renderings of astrophysical objects and processes.
His research interests include the 3D reconstruction and visualization of astrophysical nebulae based on imaging and spectroscopic data.\end{IEEEbiography}
\begin{IEEEbiography}[{\includegraphics[width=1in,height=1.25in,clip,keepaspectratio]{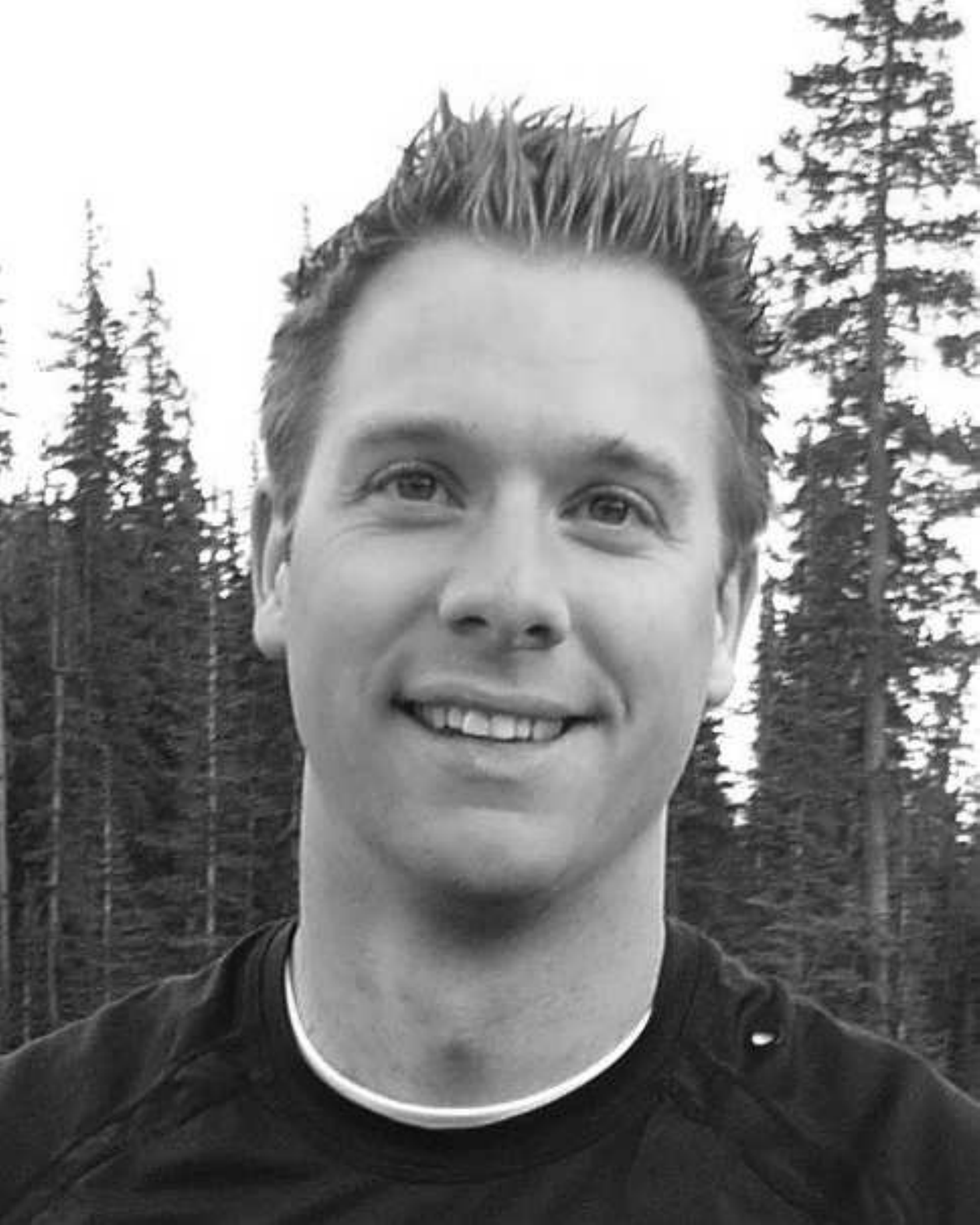}}]{Nicholas Koning} is a Ph.D. student in astrophysics at the University of Calgary in Calgary, Canada.  His research includes the development of Shape and the reconstruction of 3D morphologies of Planetary Nebulae and other astronomical objects.\end{IEEEbiography}
\begin{IEEEbiography}[{\includegraphics[width=1in,height=1.25in,clip,keepaspectratio]{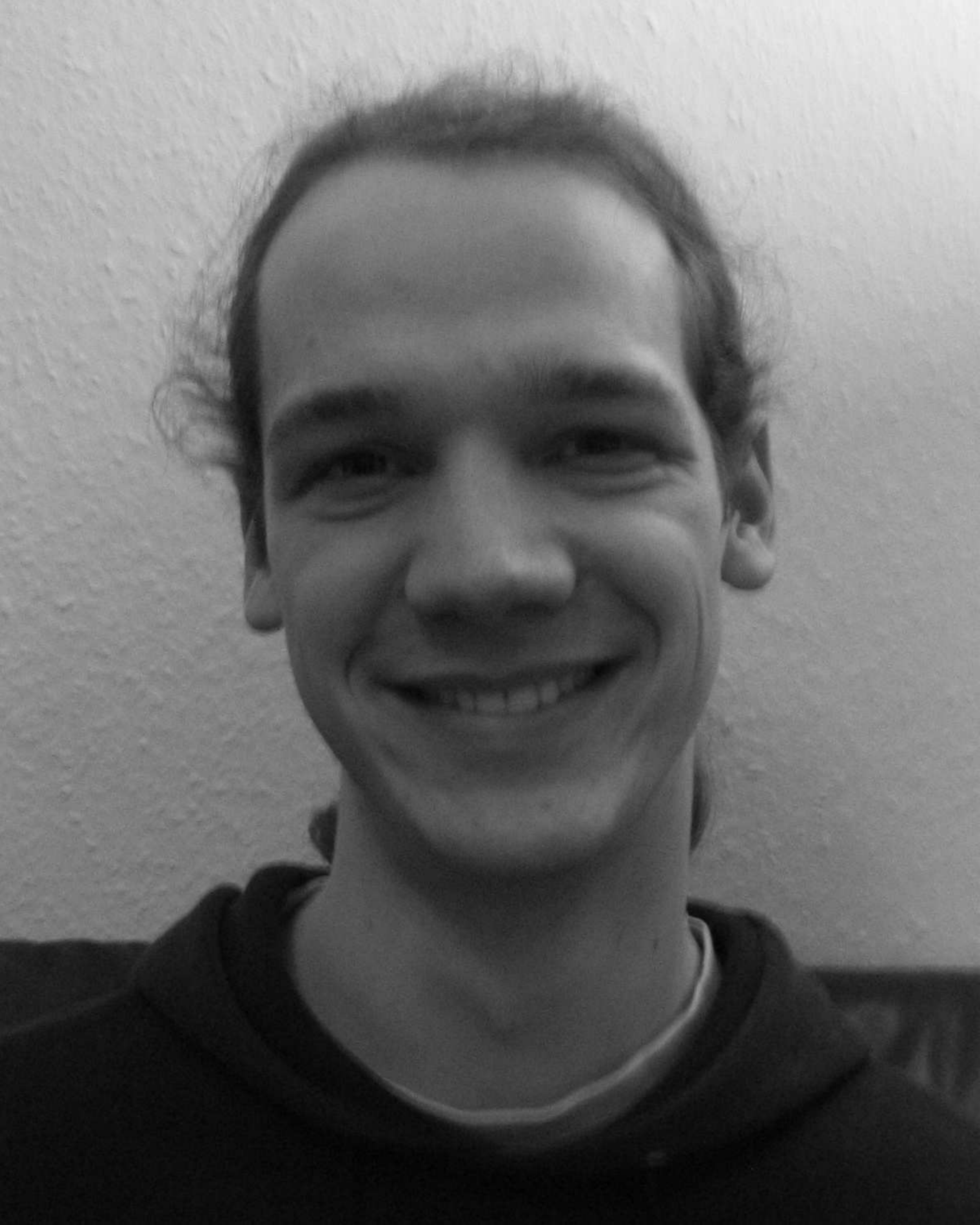}}]{Stephan Wenger} is a graduate student in physics and computer science at the University of Technology in Braunschweig, Germany. His research interests include reconstruction and visualization of astrophysical phenomena, volume visualization techniques and computational physics.\end{IEEEbiography}
\begin{IEEEbiography}[{\includegraphics[width=1in,height=1.25in,clip,keepaspectratio]{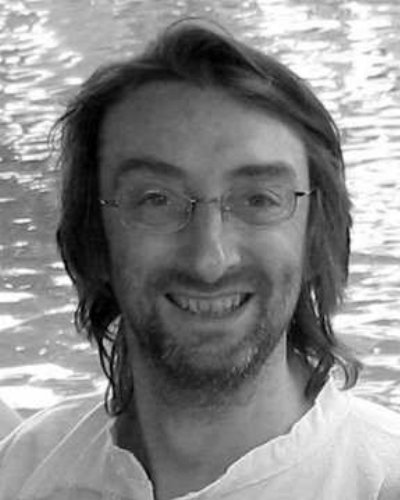}}]{Christophe Morisset} is a full-time professional astronomer working at the Institute of Astronomy of the Universidad Nacional Aut\'onoma de M\'exico in Mexico City, Mexico. He obtained his PhD in 1996 from the University Paris 7 at the Meudon Observatory. He is working on the modeling of photoionized gaseous nebulae, in particular planetary nebulae and H-II regions. He worked at the IAG (S\~{a}o Paulo, Brazil) and the IAS (Marseille, France). In the last ten years he dedicated most of its work to the development and use of 3D photoionization codes able to take into account more complex morphologies than pure spheres.\end{IEEEbiography}
\begin{IEEEbiography}[{\includegraphics[width=1in,height=1.25in,clip,keepaspectratio]{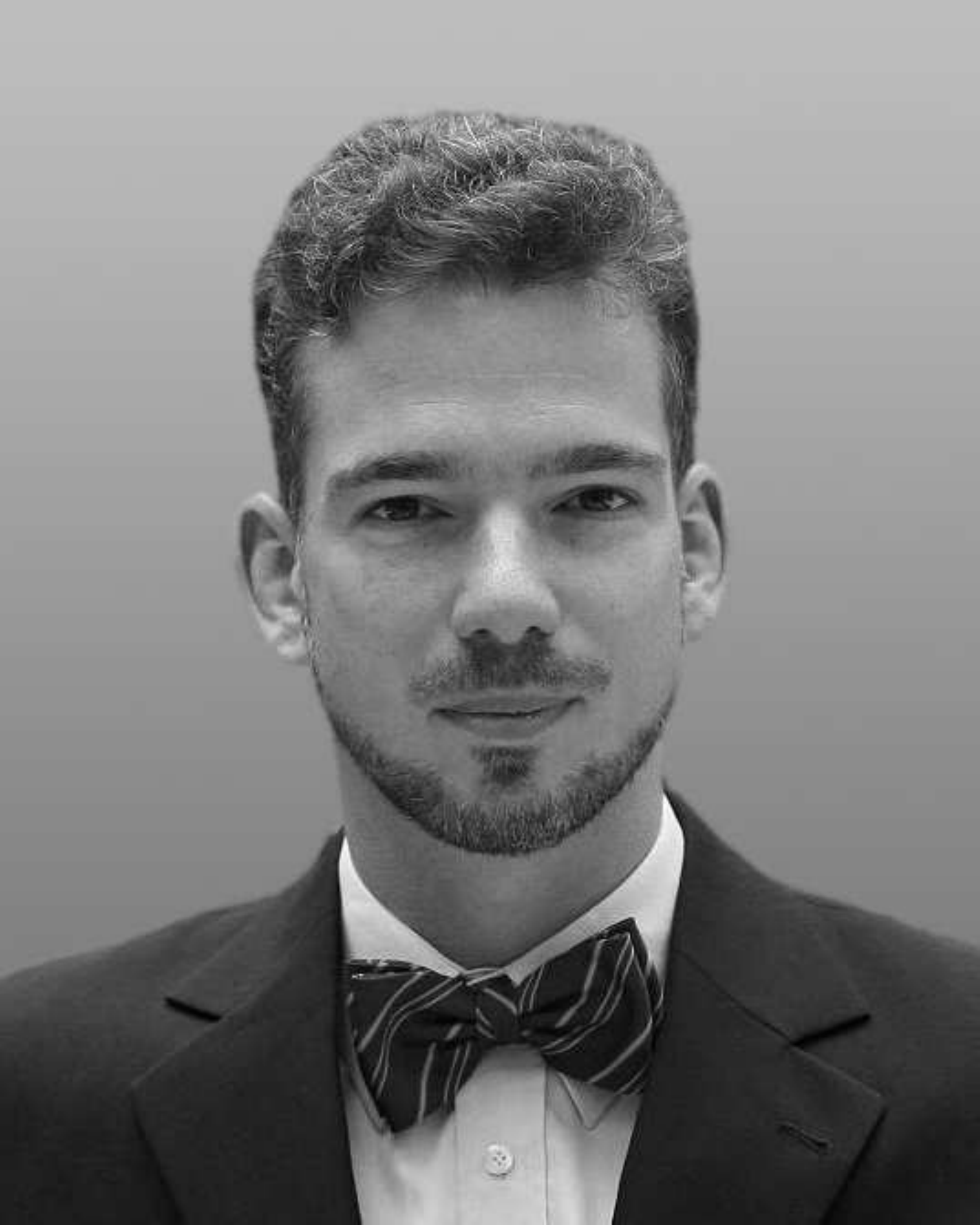}}]{Marcus Magnor} heads the Computer Graphics Lab of the Computer Science Department at Braunschweig University of Technology.
He received his BA~(1995) and MS~(1997) in physics from the University of W\"{u}rzburg and the University of New Mexico, respectively,
and his PhD~(2000) in Electrical Engineering from the Telecommunications Lab at the University of Erlangen.
For his post-graduate studies, he joined the Computer Graphics Lab at Stanford University.
In 2002, he established the Independent Research Group Graphics--Optics--Vision at the Max-Planck-Institut Informatik in Saarbr\"{u}cken.
He completed his habilitation and received his venia legendi in Computer Science from Saarland University in 2005.
His research interests meander along the visual information processing pipeline,
from image formation, acquisition, and analysis to image synthesis, display, perception, and cognition.
Ongoing research topics include image-based measuring and modeling, photo-realistic \& real-time rendering, and perception in graphics.\end{IEEEbiography}

\end{document}